\documentclass[12pt,oneside,reqno,reprint,onecolumn,nofootinbib,notitlepage,showkeys,showpacs]{revtex4-1}
\usepackage{amsfonts}
\usepackage[a4paper,left=1in,right=1in,top=1in,bottom=1in,footskip=.25in]{geometry}
\usepackage{graphicx}
\usepackage{subcaption}
\usepackage[noabbrev]{cleveref}
\usepackage{amsmath}
\usepackage{amssymb}
\usepackage{epstopdf}
\usepackage{color}
\usepackage{hyperref}

\begin{document}

\title{Evolution of Cyclic Mixmaster Universes with Non-comoving Radiation }
\date{\today }
\author{Chandrima Ganguly and John D. Barrow}
\affiliation{DAMTP, Centre for Mathematical Sciences,\\
University of Cambridge,\\
Wilberforce Rd., Cambridge CB3 0WA, U.K.}

\begin{abstract}
We study a model of a cyclic, spatially homogeneous, anisotropic, `mixmaster'
universe of Bianchi type IX, containing a radiation
field with non-comoving (`tilted' with respect to the tetrad frame of
reference) velocities and vorticity. We employ a combination of numerical
and approximate analytic methods to investigate the consequences of the
second law of thermodynamics on the evolution. We model a smooth
cycle-to-cycle evolution of the mixmaster universe, bouncing at a finite
minimum, by the device of adding a comoving `ghost' field with negative
energy density. In the absence of a cosmological constant, an increase in
entropy, injected at the start of each cycle, causes an increase in the
volume maxima, increasing approach to flatness, falling velocities and
vorticities, and growing anisotropy at the expansion maxima of successive
cycles. We find that the velocities oscillate rapidly as they evolve and
change logarithmically in time relative to the expansion volume. When the
conservation of momentum and angular momentum constraints are imposed, the
spatial components of these velocities fall to smaller values when the
entropy density increases, and vice versa. Isotropisation is found to occur
when a positive cosmological constant is added because the sequence of
oscillations ends and the dynamics expand forever, evolving towards a quasi
de Sitter asymptote with constant velocity amplitudes. The case of a single
cycle of evolution with a negative cosmological constant added is also
studied.

\end{abstract}

\maketitle

% Activate to display a given date or no date

\section{Introduction}

Cyclic models are popular alternatives to inflationary paradigms as
candidates for a viable theory of the early universe that avoids or
mitigates the singularities in the simple Friedmann universes. For this to
be a suitable theory, it must reproduce some of the successes of inflation.
One of these is to produce a high degree of isotropy at late times. In order
to discuss the process of isotropisation, we will consider the most general
spatially homogeneous closed universe with a non-comoving velocity field.
This generalises our earlier work without non-comoving velocities and will
allow us to study the behaviour of velocities and dynamics of a general
closed cyclic universe over many cycles.

The simplest cyclic universes were constructed in dust-filled or
radiation-filled closed Friedmann universes \cite{tol}. Using these simple
models, Tolman was able to show that oscillating Friedmann universes with
zero cosmological constant displayed successive cycles of increasing maximum
size and duration, see also ref. \cite{znov} for a more general result.
Later, this study was generalised to show that, when a positive cosmological
constant is included, Tolman's cycles approach flatness but always come to
an end: the dynamics ends in a state of expansion evolving towards a de
Sitter universe \cite{bdab}. If the entropy increase from cycle to cycle is
modest then his final state displays close proximity to flatness with a
slight domination by dark energy (the cosmological constant stress) over
cold dark matter (or radiation).

These isotropic models are highly idealised, especially near the initial and
final singularities, or expansion minima, of cyclic universes. We need to
generalise them by studying the shape evolution of the most general, closed,
cyclic, anisotropic universes. A start was made on this problem by
considering anisotropic Kantowski-Sachs closed universes and Bianchi type I
universes with a negative cosmological constant in ref. \cite{bdab}. This
was generalised to the closed spatially homogeneous universe with comoving
fluid velocities, of Bianchi type IX, by the authors in ref. \cite{me}. This
`mixmaster' cosmological model contains the closed Friedmann model as a
special case but introduces several new factors, including anisotropic
expansion rates (shear) and anisotropic 3-curvature, which can change sign
in the course of the evolution of the universe. This feature is
intrinsically general relativistic and these models have no Newtonian
counterparts. They allow us to study the evolution of anisotropy over a
sequence of cosmological cycles. In ref. \cite{me} we showed that this
evolution displays chaotic sensitivity to past conditions and successive
maxima of a closed universe with increasing entropy will get larger and
increasingly approach flatness, as in the isotropic case, but they will
become increasingly anisotropic at these growing maxima with respect to
their expansion anisotropy (shear) and 3-curvature anisotropy.

In ref. \cite{me}, we studied the behaviour of a Bianchi IX universe with
comoving matter velocities. Here, we extend this by introducing the last
remaining physical generalisation available to this metric -- the inclusion
of matter (or radiation) which moves with a $4$-velocity that is not
comoving with the tetrad frame. In the context of the orthonormal frame
formalism, the system contains shear, anisotropic spatial curvature, and
vorticity, \cite{matzner}. The non-comoving fluid velocities are tilted with
respect to normals to the hypersurfaces of constant density \cite{kingellis}.

There have been studies of non-comoving matter in other Bianchi types, such
as Bianchi type VII$_{0}$, such as in \cite{LukashVII, Lukashrev,
lukashnovikov, lukashstarob}, but with a focus on the fate of primordial
turbulence and the effects of collisionless particles on vortices. There has
also been a study of the problem in a radiation-filled universe in ref. \cite%
{lukash}. Here we extend these analyses, and those of cyclic universes, to
include a Bianchi IX (`mixmaster') universe with non-comoving radiation. We
also include a comoving null energy condition (NEC) violating, or `ghost',
field with $p>\rho $ and $\rho <0$ \cite{tsag, kbm}. Its inclusion is simply
a device to produce a bounce at a finite minimum of each cycle. It avoids
the evolution falling into an open interval around $t=0$ which will produce
chaotic mixmaster oscillations \cite{mis} and thus avoids the inclusion of
infinite mixmaster oscillations on approach to the expansion minima. These
oscillations are not likely to be physically relevant in classical
cosmological evolution: if the universe bounces at the Planck time ($%
t_{pl}\sim 10^{-43}s$) then only a few mixmaster oscillations are permitted
up to the present ($t_{0}$ $\sim 10^{60}t_{pl}$) because they occur in
log-log of the comoving proper time. This makes our problem more tractable from a numerical perspective. Our aim is study the effect of
non-comoving radiation in the case of a cyclic mixmaster universe with
thermal entropy growth, in the presence of both zero, positive and negative
cosmological constant.

In section 2, we set up the Einstein equations for the problem, with a brief
background to the tetrad formalism in Bianchi IX models, and give the
energy-momentum tensor of the non-comoving radiation field that we are
introducing. We derive the equations of motion and the evolution equation
for the velocities that are normalised with the appropriate power of the
energy density of radiation. In section 3, we discuss the qualitative
effects of entropy increase on the cycle to cycle evolution of closed
isotropic and anisotropic universes and identify a new effect of entropy
increase introduced by the presence of non-comoving velocities. In section
4, we provide some analytic analysis of the type IX equations with
velocities before presenting our computational solutions of the Einstein
equations with and without a cosmological constant of either sign in section
5, and give our conclusions in section 6.

\section{The setup}

\subsection{The Einstein equations}

For the purposes of studying the effect of non-comoving velocities in
anisotropic closed cyclic universes, we choose the Bianchi IX universe. In
general, when studying an $n$-dimensional spatially homogeneous, anisotropic
cosmology, we consider a group of $n$ linearly independent differential
forms which remain invariant under a group of simply transitive motions
following ref. \cite{HeckShuk,chern}, 
\begin{equation}
e_{\mu }^{a}(x^{\prime \nu })dx^{\prime \mu }=e_{\mu }^{a}(x^{\nu })dx^{\mu
},
\end{equation}%
where $x^{\prime \mu }$ and $x^{\nu }$ are the coordinates in the
transformed and the starting coordinate systems, respectively. We can then
write down an invariant metric, 
\begin{equation}
ds^{2}=\gamma _{ab}e_{\mu }^{a}(x^{\prime })e_{\nu }^{b}(x^{\prime
})dx^{\prime \mu }dx^{\prime \nu }=g_{\mu \nu }(x^{\prime })dx^{\prime \mu
}dx^{\prime \nu }.
\end{equation}%
As the line element itself remains invariant under these transformations, we
can also write 
\begin{equation}
ds^{2}=\gamma _{ab}e_{\mu }^{a}(x)e_{\nu }^{b}(x)dx^{\mu }dx^{\nu }=g_{\mu
\nu }(x)dx^{\mu }dx^{\nu }.
\end{equation}%
Considering the transformation between $x^{\prime \mu }$ and $x^{\mu }$, 
\begin{equation}
\frac{\partial x^{\prime \lambda }}{\partial x^{\mu }}=e_{a}^{\lambda
}(x_{\nu }^{\prime })e_{\mu }^{a}(x^{\rho }),
\end{equation}%
and using the fact that double differentiation must commute under
interchange of the indices $\mu $ and $\nu $, 
\begin{equation}
\frac{\partial ^{2}x^{\prime \lambda }}{\partial x^{\mu }\partial x^{\nu }}-%
\frac{\partial ^{2}x^{\prime \lambda }}{\partial x^{\nu }\partial x^{\mu }}%
=0,
\end{equation}%
we can define the following commutation relation, 
\begin{equation}
e_{d,\alpha }^{\mu }e_{c}^{\alpha }-e_{c,\alpha }^{\mu }e_{d}^{\alpha
}=C_{cd}^{f}e_{f}^{\mu }.
\end{equation}%
The $C_{cd}^{f}$ are the structure constants of the Lie algebra, and obey
the Jacobi identities: 
\begin{equation}
C_{cd}^{f}C_{fe}^{g}+C_{de}^{f}C_{fc}^{g}+C_{ec}^{f}C_{fd}^{g}=0.
\end{equation}%
We can choose to work in a system of coordinates that is more suited to our
purpose. We are dealing with spatially homogeneous cosmologies and we are
able to define coordinates on the spatial hypersurface where $t=const$ and
the comoving proper time coordinate, $t$, will just measure the distance
between parallel hypersurfaces. Thus we write the metric now as, 
\begin{equation}
ds^{2}=dt^{2}-g_{ik}dx^{i}dx^{k}.
\end{equation}%
The simply transitive group of motions that leaves the differential forms
invariant now acts on the three-spaces where $t=const$. We can then write
out the Ricci tensor components in terms of the metric: 
\begin{align}
R_{00}& =(\mathrm{ln}\sqrt{-g})\ddot{\phantom{h}}+\frac{1}{4}g^{lm}\dot{g}%
_{mk}g^{kj}\dot{g}_{jl}, \\
R_{0k}& =\frac{1}{2}g^{lm}(\dot{g}_{lm;k}-\dot{g}_{lk;m}), \\
R_{ij}& =R_{ij}^{\star }+\frac{1}{2}\ddot{g}_{ij}-\frac{1}{2}\dot{g}%
_{im}g^{mk}\dot{g}_{kj}+\frac{1}{2}\dot{g}_{ij}(\mathrm{ln}\sqrt{-g}).
\end{align}%
In our case we introduce the metric of the diagonal Bianchi IX universe, 
\begin{equation}
ds^{2}=dt^{2}-\gamma _{ab}e_{\mu }^{a}e_{\nu }^{b}dx^{\mu }dx^{\nu },
\end{equation}%
where 
\begin{equation}
\gamma _{ab}=\mathrm{diag}[a(t)^{2},b(t)^{2},c(t)^{2}].
\end{equation}

Turning our attention now to the matter sector, we introduce the
energy-momentum tensor for a perfect field: 
\begin{equation}
T_{ab}=(\rho +p)u_{a}u_{b}-p\gamma _{ab}.
\end{equation}%
The $4$-velocity of the perfect fluid with respect to our chosen tetrad
frame is 
\begin{equation}
u_{a}=(u_{0},u_{1},u_{2},u_{3}).
\end{equation}%
The relations of the components of the velocities with respect to the
universe frame are given by, 
\begin{equation}
u_{a}=e_{a}^{\mu }\bar{u}_{\mu },
\end{equation}%
where the $\bar{u}_{\mu }$ are the components of the $4$-velocity of the
fluid with respect to the universe frame. We shall be working with the $4$%
-velocity in the tetrad frame for consistency with the Ricci tensor, which
is also written in the tetrad frame for the purposes of this computation.
The components of this $4$-velocity obey the normalisation, 
\begin{equation}
u_{0}^{2}-u_{1}^{2}-u_{2}^{2}-u_{3}^{2}=1.
\end{equation}%
Referring to \cite{matzner}, we find the following conditions on the
energy-momentum tensor. The fluid vorticity $\omega _{ab}$ is zero if and
only if the spatial velocity components $u_{i}$ are zero. Thus, for the
general case of non-comoving fields, we do indeed have vorticity in our
system. Thus, with reference to the orthonormal frame formalism, we have
non-zero shear $\sigma _{ab}$, the curvature variables $n_{ab}$ as well as
vorticity $\omega _{ab}$. \\
For the purposes of our computation, we consider
non-interacting perfect fluids, with ideal equation of state, 
\begin{equation}
p=(\gamma -1)\rho ,
\end{equation}%
and we can add their energy-momentum tensors together in the usual way. In
our system, we include radiation with $\gamma \equiv \gamma _{r}=4/3$ and an
ultra-stiff comoving `ghost' field with equation of state $\gamma \equiv
\gamma _{g}=5,$a value chosen simply for convenience in effecting a bounce.
The densities and the pressures of the radiation and the `ghost' field are
given by $\rho _{r}$, $p_{r}$ and $\rho _{g}$ and $p_{g}$. The radiation
field has velocities which are not comoving in the tetrad frame of
reference. We normalise the $4$-velocity of the radiation field so that the
normalised velocity components are related to the velocity vector by $%
v_{a}=(\rho _{r}+p_{r})^{1/\gamma -1/2}u_{a}$, and denote the normalised
velocity vector by 
\begin{equation}
\mathbf{v}=\left( v_{0},v_{1},v_{2},v_{3}\right) 
\end{equation}%
In our case, for black body radiation, $\gamma _{r}=4/3$ and the normalised
velocity components are therefore given by $v_{a}=(\rho
_{r}+p_{r})^{1/4}u_{a}$. Considering energy-momentum conservation in the
tetrad frame, we get the conservation of particle current, 
\begin{equation}
\frac{1}{\sqrt{-g}}\frac{\partial }{\partial x^{i}}(\sqrt{-g}su^{i})=0,
\label{eq:current_conservation}
\end{equation}%
where $s$ is the entropy density. For radiation, $s\propto \rho ^{3/4}$,
this yields the conservation law, 
\begin{equation}
v_{0}^{2}a^{2}b^{2}c^{2}(\rho _{r}+p_{r})=const\equiv L^{3},
\label{eq:v0_constraint}
\end{equation}%
where we have labelled the constant $L^{3}$ for consistency with reference 
\cite{lukash}.

The second constraint equation for the components of velocity of the
radiation field is 
\begin{equation}
v_{1}^{2}+v_{2}^{2}+v_{3}^{2}=L\delta  \label{eq:vel_constraint}
\end{equation}%
The constant $L$ has the dimensions of length and the constant $\delta $ is
dimensionless. Close to isotropy, when the spatial components of the
velocity $4$-vector are negligible, we have $\delta \ll 1$. For the case of small velocities in a near- Friedmann radiation-dominated universe, we see
that their spatial components are constant. For the
dust-dominated universe, the spatial components of the velocities fall as $%
1/a$ where $a(t)$ is the scale factor of the Friedmann universe and $t$ is
the comoving proper time.

We have a further hydrodynamic equation of motion ($\nabla ^{a}T_{ab}=0$),
and 4-velocity normalisation \cite{landau} to employ in what follows:

\begin{eqnarray}
(p+\rho )u^{k}\left( \frac{\partial u_{i}}{\partial x^{k}}-\frac{1}{2}u^{l}%
\frac{\partial g_{kl}}{\partial x^{i}}\right) &=&-\frac{\partial p}{\partial
x^{i}}-u_{i}u^{k}\frac{\partial p}{\partial x^{k}} \\
u_{i}u^{i} &=&1
\end{eqnarray}

\subsection{Non-comoving velocities in a Bianchi type IX universe}

We now ask what happens in an anisotropic, spatially homogeneous universes,
with scale factors $a,b,c$, when there are non-comoving velocities and
vorticities. Suppose we first take the background expansion of the scale
factors to have the same form as in the type IX universe without
non-comoving velocities, that we studied in ref. \cite{me}. In our earlier
study without velocities we found a long period of evolution during the
radiation era (before the curvature creates a slow-down of the expansion
near the volume maximum) where,far
from the expansion maximum, the scale factors evolve to a good
approximation in a quasi-axisymmetric manner, as

\begin{equation}
a(t)=a_{0}t^{1/2}[\ln (t)]^{-1/2},b(t)=b_{0}t^{1/2}[\ln
(t)]^{-1/2},c(t)=c_{0}t^{1/2}\ln (t).  \label{axi}
\end{equation}

Note that the volume, $abc\propto t^{3/2}$, evolves like the Friedmann model 
\cite{lukash}. The logarithmic corrections are familiar in the study of
anisotropic universes with anisotropic 3-curvatures, trace-free radiation
stresses in the presence of isotropic radiation, or long-wavelength
gravitational waves \cite{skew}. They reflect the presence of a zero
eigenvalue when we perturb around the shear variables around the isotropic
model whereas the volume has a negative real eigenvalue.\\
 More generally, the
effects of the velocities in the type IX radiation universe can be treated
as test motions on an expanding radiation background governed by equations %
\eqref{eq:current_conservation} and \eqref{eq:v0_constraint}:

\begin{subequations}
\begin{align}
&a(t) =a_{0}t^{1/2}\ln ^{\lambda }(t),b(t)=b_{0}t^{1/2}\ln ^{\mu
}(t),c(t)=c_{0}t^{1/2}\ln ^{\nu }(t),  \label{nonaxi1} \\
&\lambda +\mu +\nu=0,\text{and }\lambda ,\mu ,\nu \text{ constants,}
\label{nonaxi2} \\
&abc \propto t^{3/2},  \label{nonaxi3}
\end{align}

where (\ref{nonaxi1}) reduces to the particular case (\ref{axi}) when $%
\lambda =\mu =-1/2$ and $\nu=1$.

Ignoring spatial gradients with respect to time variations, and taking the
diagonal scale factors to be $a(t),b(t),$and $c(t)$, for a
radiation-dominated universe ($p=\rho /3$), these equations specialise to 
\cite{landau}

\end{subequations}
\begin{eqnarray}
abcu_{0}\rho ^{3/4} &=&t^{3/2}u_{0}\rho ^{3/4}=\mathrm{constant,}  \label{1}
\\
u_{\alpha }\rho ^{1/4} &=&\mathrm{constant,}\alpha =1,2,3.  \label{2}
\end{eqnarray}

If we solve them as $t\rightarrow \infty $ with $\lambda <\mu <\nu ,$ then
the dominant component of $u^{\alpha }$ is $u^{1}=u_{1}/a^{2},$which gives $%
u_{0}^{2}\simeq u_{1}u^{1}=(u_{1})^{2}t^{-1}\ln ^{-2\lambda }(t)$, and we
get the dominant late-time behaviours from (\ref{1})-(\ref{2}):

\begin{eqnarray}
\rho &\propto &\frac{\ln ^{2\lambda }(t)}{t^{2}},u_{1}u^{1}\propto \frac{1}{%
\ln ^{3\lambda }(t)},  \label{3} \\
T_{1}^{1} &\simeq &\rho u_{1}u^{1}\propto \frac{1}{t^{2}\ln ^{\lambda }(t)}%
\propto T_{0}^{0}  \label{4} \\
T_{2}^{2} &\simeq &\rho u_{2}u^{2}\propto \frac{\ln ^{\lambda -2\mu }(t)}{%
t^{2}}  \label{5} \\
T_{3}^{3} &\simeq &\rho u_{3}u^{3}\propto \frac{\ln ^{\lambda -2\nu }(t)}{%
t^{2}}  \label{6}
\end{eqnarray}

The corrections to the case with comoving velocities and zero vorticity are
therefore only logarithmic in time during the radiation era. The scalar
3-velocity, has dominant asymptotic form%
\begin{equation}
V\equiv \sqrt{u_{\alpha }u^{\alpha }}\simeq \ln ^{-3\lambda }(t).
\end{equation}

For the quasi-axisymmetric radiation-dominated phase of the type IX
evolution, we take $\lambda =\mu =-1/2$ and $\nu =1$ and we see that the
stresses induced by the velocities grow logarithmically in time compared to
the other terms in the field equations (of order $O(1/t^{2}$)) present when
the velocities are comoving. We see that the diagonal stress-tensor components $%
T_{0}^{0}\propto T_{1}^{1}\propto T_{2}^{2}\propto t^{-2}\ln ^{1/2}(t)$ fall
off slower than $t^{-2}$ as $t\rightarrow \infty ,$ while $T_{3}^{3}\propto
t^{-2}\ln ^{-3/2}(t)$ falls off faster than $t^{-2}$. We can see explicitly
that the 3-velocity, $V$, is expected to grow as $\ln ^{3/2}(t)$ in our
approximation, which holds so long as the velocities are small enough for
the perturbations not to disrupt the assumed (velocity-free) metric
evolution \eqref{nonaxi1} and we are far from the expansion maximum. If
there is an expansion maximum, then these asymptotic forms will be cut off
when the approximate solution \eqref{axi} breaks down and we need a
numerical analysis to determine the detailed evolution in this regime, and
from cycle to cycle. However, we expect the presence of non-comoving
velocities to introduce changes to the analysis that was made for type IX
cyclic universes in our work. \cite{me}.

If we repeat this analysis in an isotropic de Sitter background with
late-time scale factor evolution approaching $a=b=c=e^{H_{o}t}$ before the
volume maximum, then the asymptotic behaviour of radiation is $u_{\alpha
}u^{\alpha }=$ const., $u_{0}=$ const.$,$ $\rho _{r}\propto e^{-4H_{0}t}$,
and the new terms induced in the field equations by the non-comoving
velocities do not grow at late times. However, we note that the velocities
produce a constant tilt relative to the normals to the surfaces of
homogeneity and the asymptotic form at late times approaches de Sitter with
a constant velocity field tilt (as also is seen in ref. \cite{star}). In
general, when the cosmological constant, $\Lambda \equiv 3H_{0}^{2}$, is
positive it will end the sequence of increasing oscillations in a cyclic
closed universe, no matter how small its value, because the size of the
universe will eventually become large enough for $\Lambda $ to dominate
before a maximum is reached in some future cycle \cite{bdab}.

\subsection{Equations of motion}

In the type IX universe, the evolution equations for the velocities are as
follows (where overdot is $d/dt$): 
\begin{equation}
\dot{v}_{1}+\frac{v_{3}v_{2}}{v_{0}}\left( \frac{1}{c^{2}}-\frac{1}{b^{2}}%
\right) \left( 1+2L^{3}w^{-1/2}\frac{a^{4}-b^{2}c^{2}}{%
(a^{2}-b^{2})(c^{2}-a^{2})}\right) =0,
\end{equation}%
\label{eq:vel1} 
\begin{equation}
\dot{v}_{2}+\frac{v_{1}v_{3}}{v_{0}}\left( \frac{1}{a^{2}}-\frac{1}{c^{2}}%
\right) \left( 1+2L^{3}w^{-1/2}\frac{b^{4}-c^{2}a^{2}}{%
(b^{2}-c^{2})(a^{2}-b^{2})}\right) =0,
\end{equation}%
\label{eq:vel2} 
\begin{equation}
\dot{v}_{3}+\frac{v_{2}v_{1}}{v_{0}}\left( \frac{1}{b^{2}}-\frac{1}{a^{2}}%
\right) \left( 1+2L^{3}w^{-1/2}\frac{c^{4}-a^{2}b^{2}}{%
(c^{2}-a^{2})(b^{2}-c^{2})}\right) =0,
\end{equation}%
\label{eq:vel3} where $w\equiv (\rho _{r}+p_{r})$. The evolution equations
for the scale factors in cosmological time then become,

\begin{align}
& (\mathrm{ln}a)\ddot{\phantom{h}}+3H(\mathrm{ln}a)\dot{\phantom{h}}+\frac{1%
}{2}\left( \frac{a^{2}}{b^{2}c^{2}}-\frac{b^{2}}{c^{2}a^{2}}-\frac{c^{2}}{%
a^{2}b^{2}}\right) +\frac{1}{a^{2}}+2L^{3}\left( \frac{a^{2}+c^{2}}{%
b^{2}(c^{2}-a^{2})^{3}}v_{2}^{2}-\frac{a^{2}+b^{2}}{c^{2}(a^{2}-b^{2})^{3}}%
v_{3}^{2}\right)  \\
& =2\left( w^{1/2}\frac{v_{1}^{2}}{a^{2}}+\frac{\rho _{r}-p_{r}}{2}-\frac{%
\rho _{g}+p_{g}}{2}\right)   \notag
\end{align}%
\begin{align}
& (\mathrm{ln}b)\ddot{\phantom{h}}+3H(\mathrm{ln}b)\dot{\phantom{h}}+\frac{1%
}{2}\left( \frac{b^{2}}{c^{2}a^{2}}-\frac{a^{2}}{b^{2}c^{2}}-\frac{c^{2}}{%
a^{2}b^{2}}\right) +\frac{1}{b^{2}}+2L^{3}\left( \frac{b^{2}+a^{2}}{%
c^{2}(b^{2}-a^{2})^{3}}v_{3}^{2}-\frac{b^{2}+c^{2}}{a^{2}(b^{2}-c^{2})^{3}}%
v_{1}^{2}\right)  \\
& =2\left( w^{1/2}\frac{v_{2}^{2}}{b^{2}}+\frac{\rho _{r}-p_{r}}{2}-\frac{%
\rho _{g}+p_{g}}{2}\right)   \notag
\end{align}%
\begin{align}
& (\mathrm{ln}c)\ddot{\phantom{h}}+3H(\mathrm{ln}c)\dot{\phantom{h}}+\frac{1%
}{2}\left( \frac{c^{2}}{a^{2}b^{2}}-\frac{a^{2}}{b^{2}c^{2}}-\frac{b^{2}}{%
a^{2}c^{2}}\right) +\frac{1}{c^{2}}+2L^{3}\left( \frac{c^{2}+b^{2}}{%
a^{2}(b^{2}-c^{2})^{3}}v_{1}^{2}-\frac{c^{2}+a^{2}}{b^{2}(c^{2}-a^{2})^{3}}%
v_{2}^{2}\right)  \\
& =2\left( w^{1/2}\frac{v_{3}^{2}}{c^{2}}+\frac{\rho _{r}-p_{r}}{2}-\frac{%
\rho _{g}+p_{g}}{2}\right)   \notag
\end{align}%
These equations include a comoving ghost field ($\rho_g$) and the non-comoving
radiation field ($\rho_r$). We give an approximate analysis of the solutions to
these equations during the radiation era, far from expansion minima and
maxima in the Appendix. We show there that the velocity components are
constant up to logarithmic oscillatory factors during the era when the
expansion dynamics are well approximated by \ref{nonaxi1}.

\section{Introducing entropy increase}

We want to investigate the effect of the non-comoving velocities on a
closed cyclic type IX universe when its radiation entropy increases from
cycle to cycle, mirroring Tolman's classic analysis \cite{tol}. The
radiation entropy density is $s\propto \rho ^{3/4}$. As in the earlier
analysis made in ref. \cite{me}, we first consider a closed Bianchi IX
universe containing radiation, dust, and a ghost field but no cosmological
constant. The ghost field has negative density and is dominant when the
singularity is approached but dynamically irrelevant far from the initial
and final singularities in each large cycle. It is included only to create a
bounce at finite volume. This avoids evolution into the open interval of
time around a curvature singularity at $t=0$ during which the dynamics will
be chaotic \cite{bkl, mis, JBIX, chern}. For realistic choices of $%
T_{1}\approx 10^{-43}s$ as the start of classical cosmology, there will be
less than about $12$ Mixmaster oscillations even if they continued all the
way from $T_{1}$ up to the present day \cite{znov, DLN2}. This is because
the overall expansion scale changes rapidly with the number of scale factor
oscillations, which occur in log-log time.

%%%%%individual and volume scale factors for entropy increase%%%%

\begin{figure}[tbp]
\caption{Evolution of (a) the volume scale factors, and (b) the individual scale
factors (left to right) with the increase in entropy with time $t$
in a Bianchi IX universe where the radiation is not comoving with the tetrad
frame, as well as a comoving dust field, and a comoving ghost field to facilitate the bounce.
The blue starred, red dotted, and green lines correspond to the
principal values of the $3$-metric in the tetrad frame, scale factors $a(t)$, $b(t)$ and $c(t)$ respectively. }%
\centering\hfill \break 
\begin{minipage}{0.48\textwidth}
\includegraphics[width=1.1\linewidth, height=0.28\textheight]{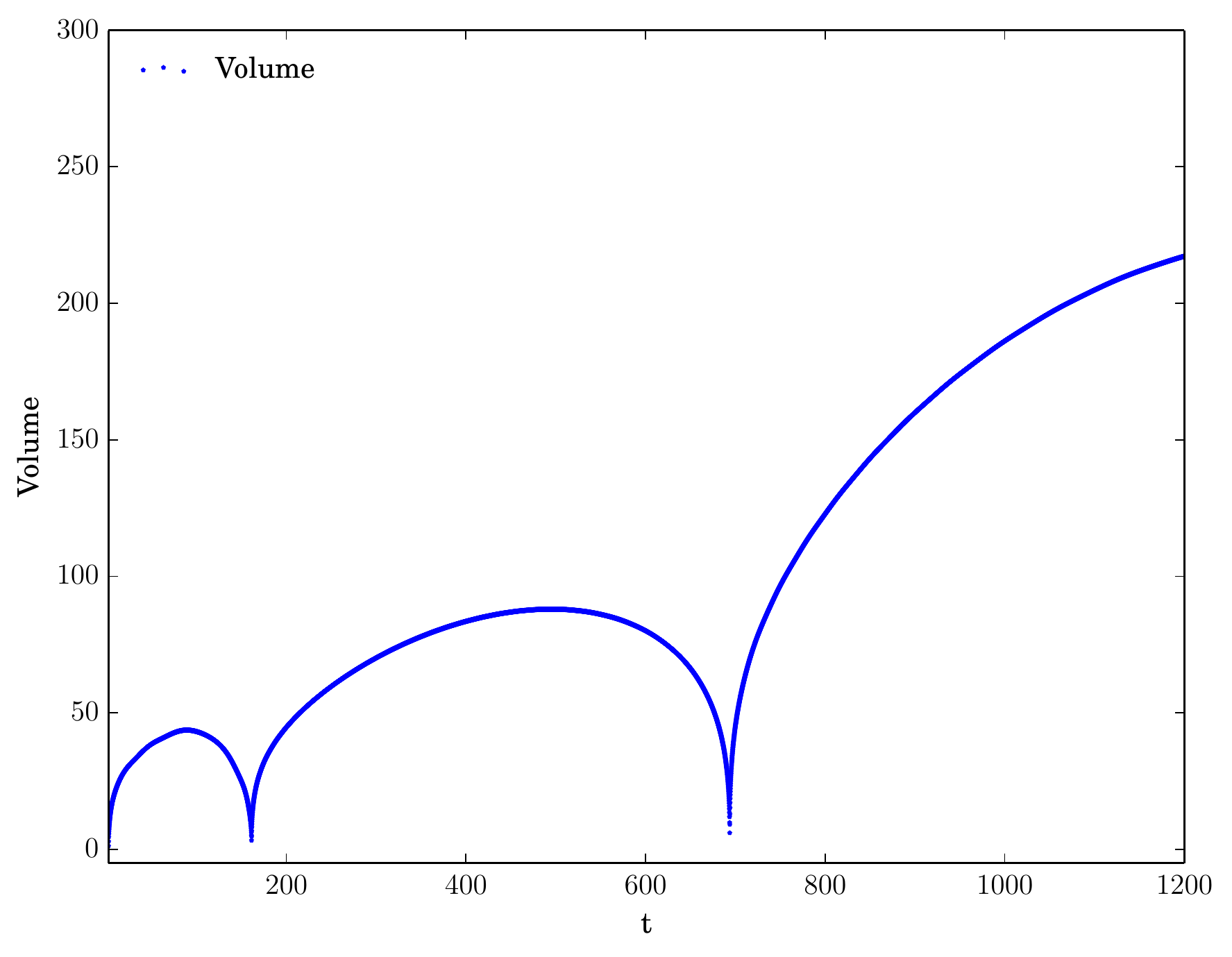}
\subcaption{\label{fig:entropyvolume}}
\end{minipage}\hfill 
\begin{minipage}{0.45\textwidth}
\includegraphics[width=1.1\linewidth,height=0.28\textheight]{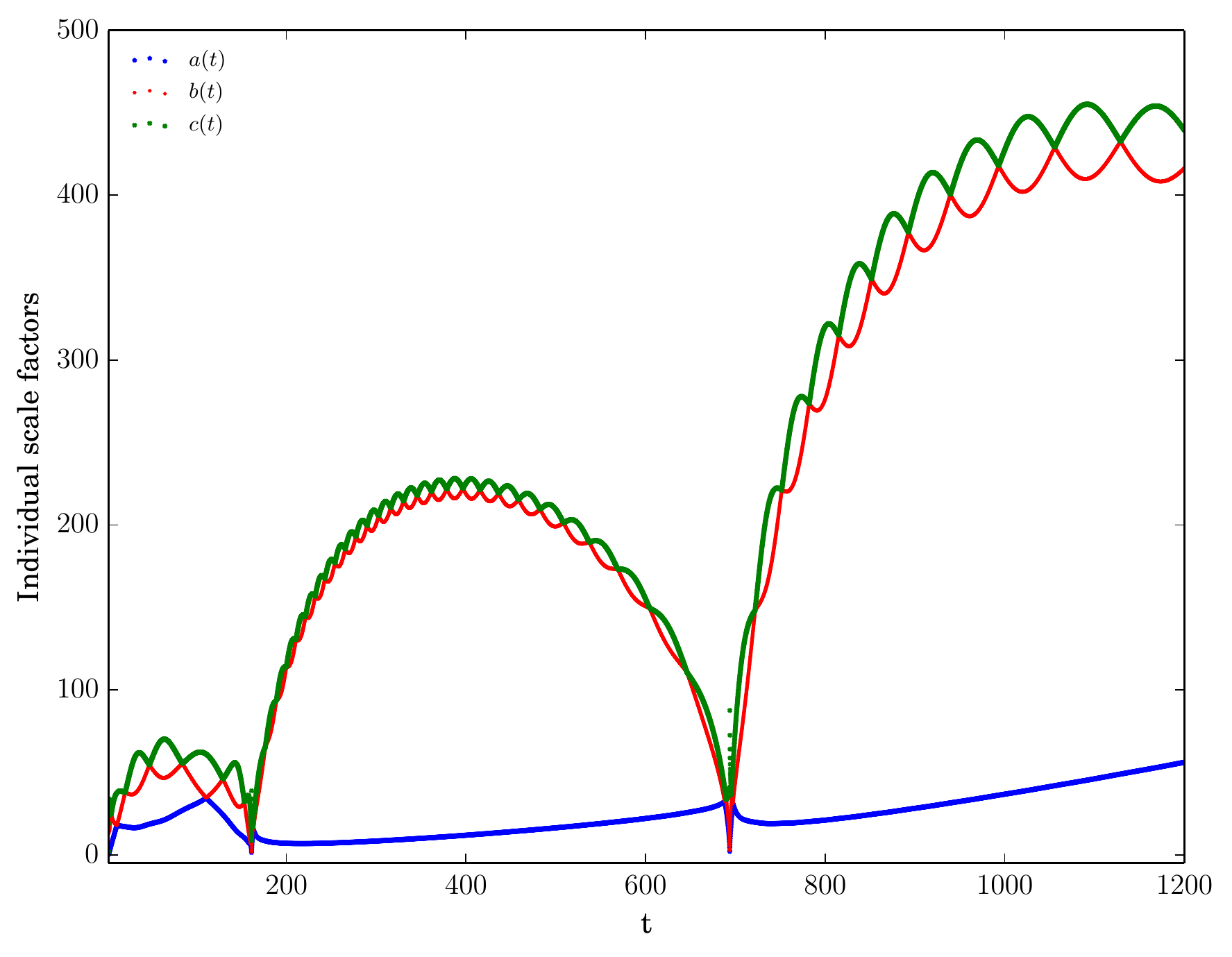}
\subcaption{\label{fig:entropyindividual}}
\end{minipage}
\end{figure}

\subsection{Effects of entropy increase}

The effects of an increase in entropy from cycle to cycle of an isotropic
oscillating closed universe were first considered by Tolman \cite{tol}. He
showed that there would be an increase in expansion volume maxima and cycle
length from cycle to cycle as a consequence of the second law of
thermodynamics. The total energy of the universe is zero in each cycle and
successive oscillations drive the universe closer and closer to flatness.

If the dynamics are allowed to be anisotropic then we showed that, with $%
\Lambda =0$, increasing entropy leads to the increase of volume maxima and
cycle length in successive cycles but the anisotropy grows from cycle to
cycle in a manner that displays sensitive dependence on `initial'
conditions. We investigated this development in the context of the Bianchi
type IX universe with comoving fluid velocities-- the most general closed
spatially homogeneous universe containing an isotropic FLRW universe as a particular case 
\cite{me}. The addition of $\Lambda >0$ eventually terminates these
oscillations, as in the isotropically expanding case.

In this paper we add an extra generalisation -- the addition of non-comoving
velocities to the most general anisotropic closed universe evolution with
entropy increase. According to \eqref{eq:current_conservation}, the entropy
increase from cycle to cycle should lead to a new effect: the reduction of
the velocity from cycle to cycle. However, it is important to keep in mind
that the constant on the right hand side of the conservation equation
resulting from \eqref{eq:current_conservation}, that is equation %
\eqref{eq:vel_constraint}, does not remain constant from cycle to cycle.
Close to isotropy, the energy density $\rho \sim L^{3}(abc)^{-4/3}$, and the
entropy density $s\propto \rho ^{3/4}$ for radiation. Increasing the entropy
density from cycle to cycle, means that $L$ remains constant only per cycle
but jumps to a higher value in the next cycle. Thus, the constraint equation %
\eqref{eq:vel_constraint} is valid in each cycle with the right hand side
being equal to a new, larger constant in subsequent cycles if there is
entropy increase. A way of modelling this problem is to ensure that the
constraint is imposed simultaneously with the injection of entropy at each
minima. Thus, if we increase the entropy, or in our case the energy density
(as $s\propto \rho ^{3/4}$) by a factor $\Delta $, then the normalised
velocities $v_{i}=\rho ^{1/4}u_{i}$ must be multiplied by a factor $\Delta
^{-1/4}$ to keep the constraint equation \eqref{eq:vel_constraint}
unchanged. Thus we see that when the entropy increases, the velocities
decrease as the evolution proceeds from cycle to cycle in accord with the
second law of thermodynamics.

\begin{figure}[tbp]
\caption{Evolution of the squares of velocities of non-comoving radiation with the increase in entropy with time $t$
in a Bianchi IX universe containing non-comoving radiation, as well as comoving
dust and the ghost fields, the latter to facilitate the bounce. The velocity constraint \eqref{eq:vel_constraint} has been imposed. An increase in entropy(energy density) causes a decrease in the velocities and vice versa. Where necessary in the last two figures, the figure has been magnified to capture the rapidly oscillating features of the plot. From left, clockwise, the entropy density ($s\propto \rho ^{3/4}$),the square of the spatial components of the velocities, $u_1^2, u_3^2$ and $u_2^2$ are shown.}%
\centering\hfill \break 
\begin{minipage}{0.45\textwidth}
\includegraphics[width=1.0\linewidth,height=0.28\textheight]{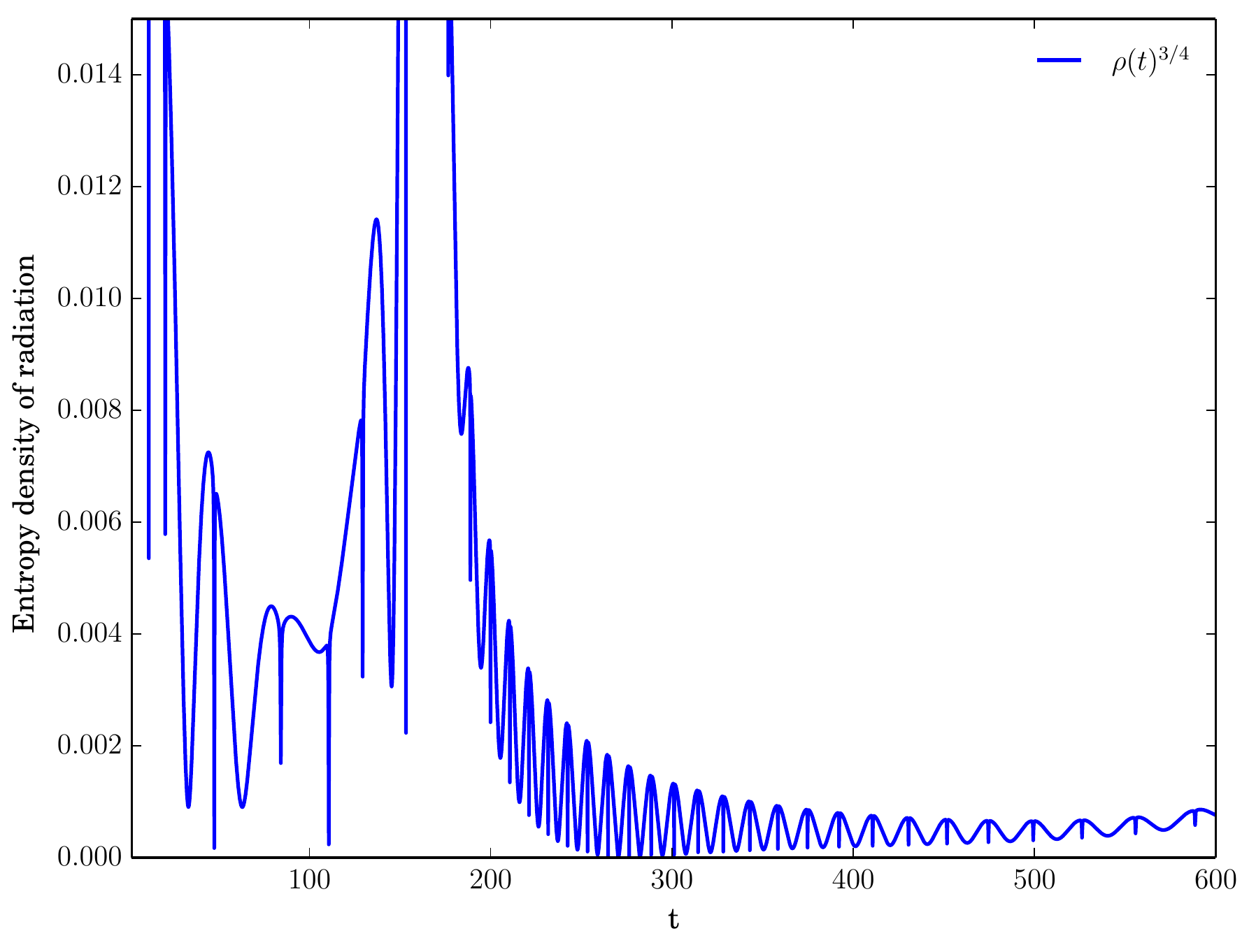}
\subcaption{\label{fig:densityconstraint}}
\end{minipage}%
\begin{minipage}{0.48\textwidth}
\includegraphics[width=1.1\linewidth, height=0.28\textheight]{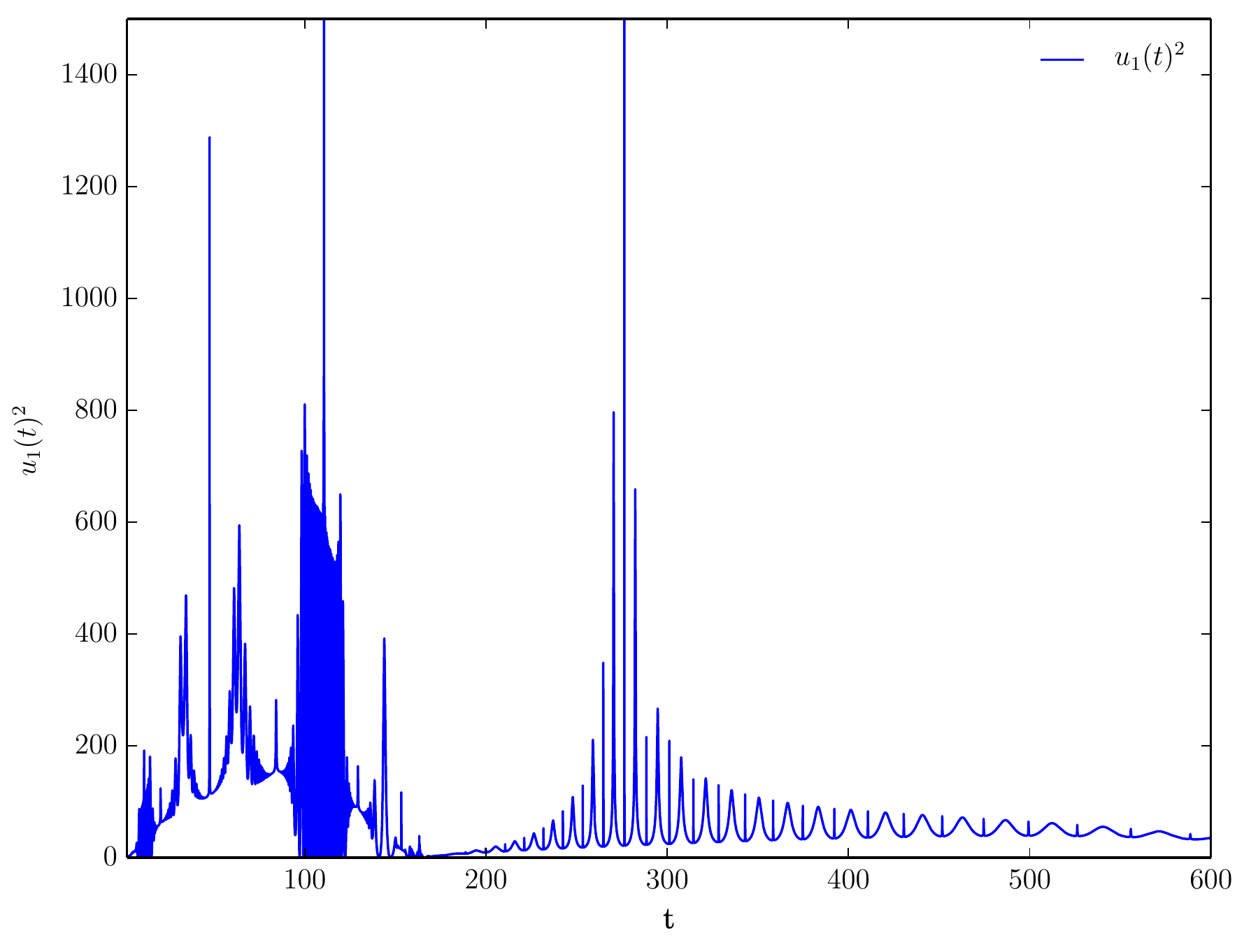}
\subcaption{\label{fig:velconstraintu1}}
\end{minipage}\hfill 
\begin{minipage}{0.45\textwidth}
\includegraphics[width=1.0\linewidth,height=0.28\textheight]{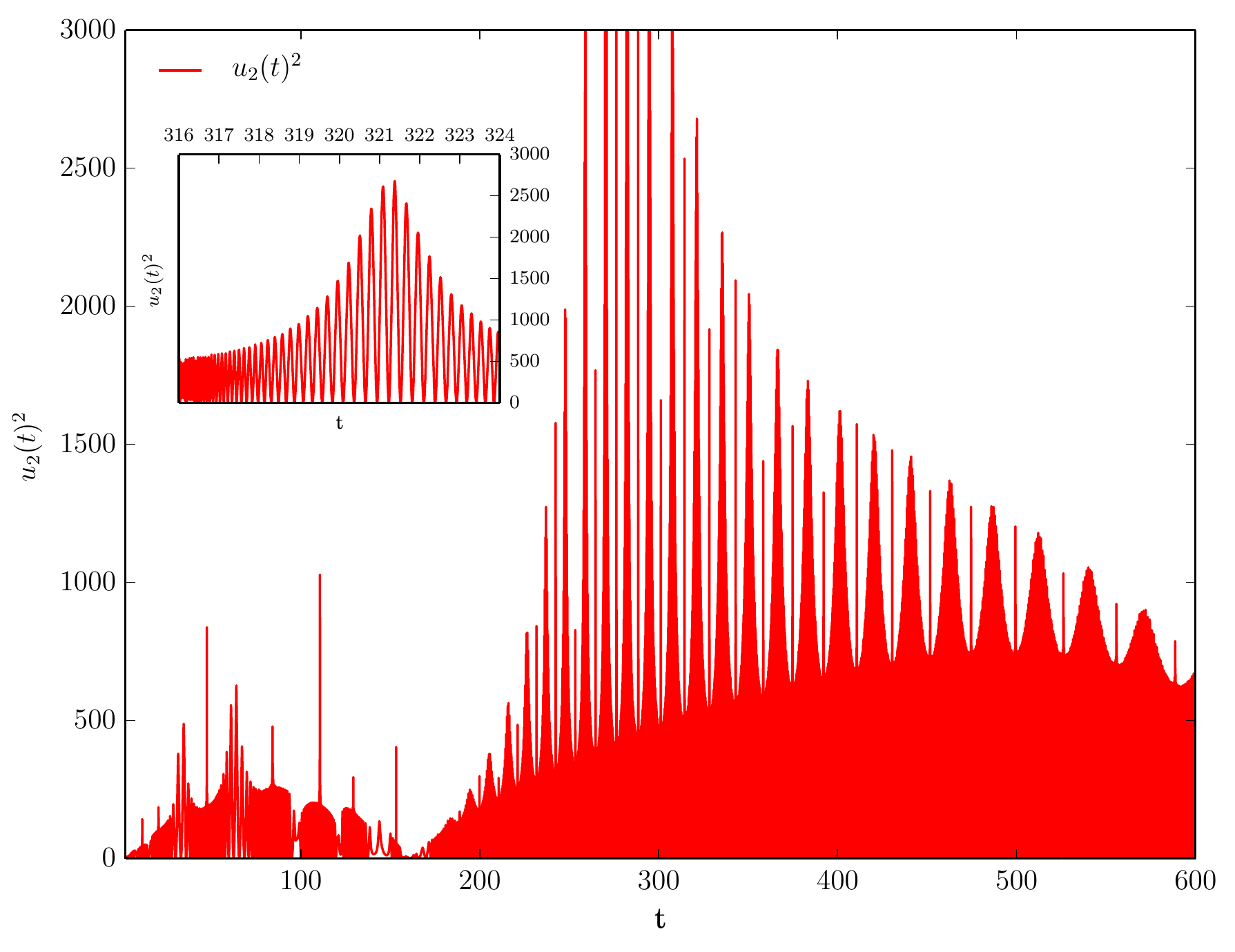}
\subcaption{\label{fig:velconstraintu2}}
\end{minipage}%
\begin{minipage}{0.45\textwidth}
\includegraphics[width=1.1\linewidth,height=0.28\textheight]{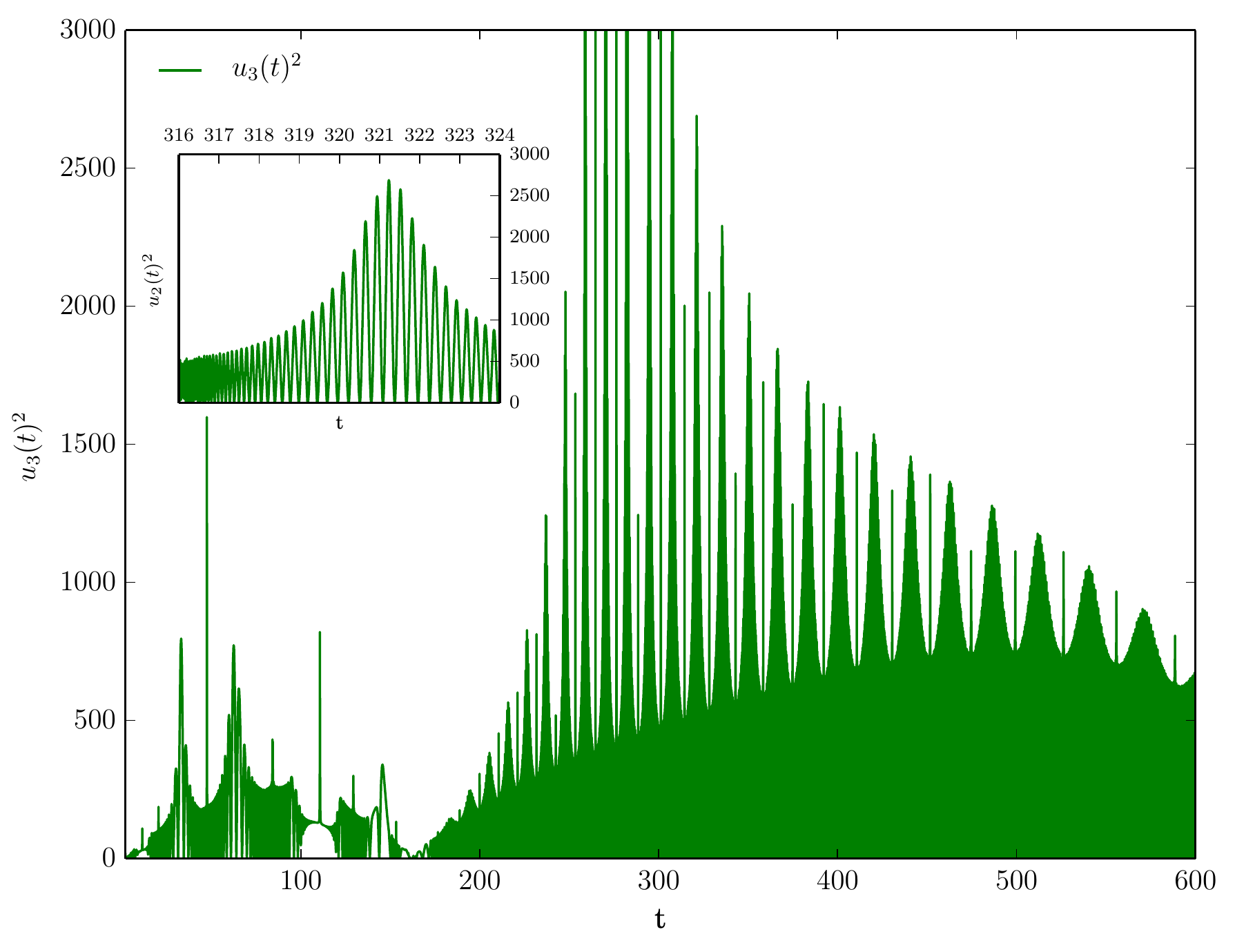}
\subcaption{\label{fig:velconstraintu3}}
\end{minipage}
\end{figure}

The sum of the square of the normalised velocities, $(\rho
+p)^{1/4}u_{\alpha }$, oscillates initially but eventually settles down to a
nearly constant value with small oscillations around this value even as
oscillations proceed to higher and higher expansion maxima. In Figure \ref%
{fig:velsqsum}, we show the constancy of this sum over one cycle. We have
modelled the effects of radiation entropy, $s$, increase during a cycle of a
closed universe by creating a sudden entropy increase at the start of each
cycle\footnote{
We assume that the additional radiation entropy is at rest relative to the
comoving frame so that we are not adding angular momentum. The situation is
analogous to the effect of quantum created particle at the Planck epoch on
vortical motions, where the increase in inertia of created particles causes
velocities to drop \cite{JB}}. This produces the increase in the expansion
maximum of successive cycles, first discovered by Tolman \cite{tol}

\begin{figure}
\includegraphics[scale=0.4]{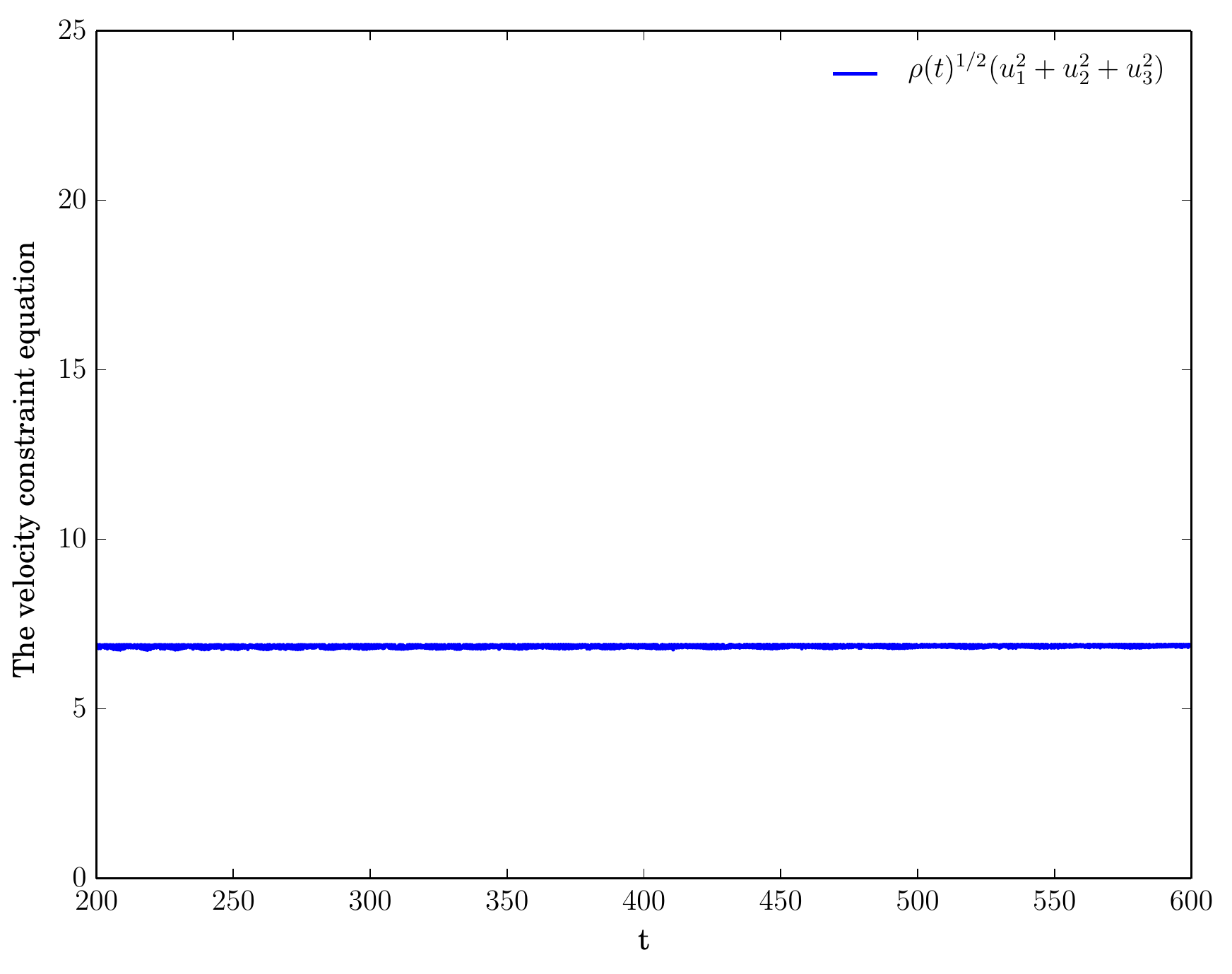} 
\caption{The evolution of
\eqref{eq:vel_constraint}, the velocity constraint equation, over one cycle} %
\label{fig:velsqsum}
\end{figure}

We identify a new feature of isotropic, oscillating radiation universes: any
non-comoving velocities and vorticities will diminish from cycle to cycle as
the expansion maxima increase and flatness is approached in accord with the
second law of thermodynamics. For the anisotropic case, the overall trend in
velocity evolution is oscillatory and is made more complicated. This is because we have
shown that flatness is approached with an increase in expansion maxima and
the inclusion of non-comoving velocities changes the dependence of the
energy density and hence of the entropy on the scale factors from the isotropic
case (and the anisotropic case in the absence of these non-comoving
velocities) \cite{me}. Thus we can only observe a increase/decrease in the
velocities with a corresponding decrease/increase in the entropy. Aside from
this effect, the evolutionary impact of the non-comoving velocities on the
evolution in a cyclic radiation universe found in case with comoving
velocities is only asymptotically logarithmic in time \cite{me}.
%%%%%shear and 3-curvature for entropy increase%%%%

\begin{figure}[tbp]
\caption{Evolution of (a) the $3$-curvature and (b)  the shear with the increase of entropy with time $t$ in a Bianchi IX universe where the radiation is not comoving with the
tetrad frame, also containing a comoving dust field and a comoving ghost field to facilitate the
bounce. }\centering\hfill \break 
\begin{minipage}{0.48\textwidth}
\includegraphics[width=1.0\linewidth, height=0.20\textheight]{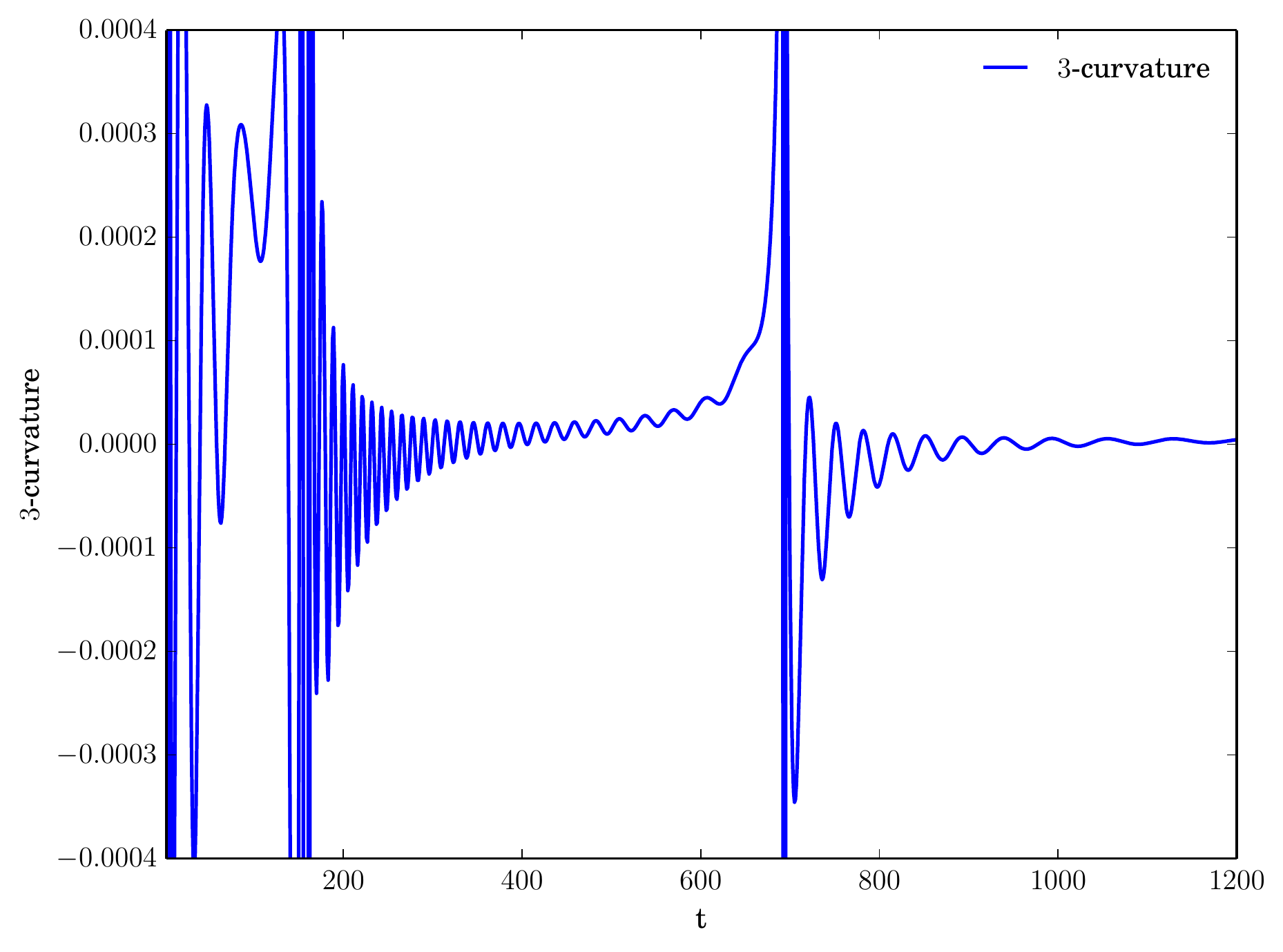}
\subcaption{\label{fig:entropycurvature}}
\end{minipage}\hfill 
\begin{minipage}{0.45\textwidth}
\includegraphics[width=1.0\linewidth,height=0.20\textheight]{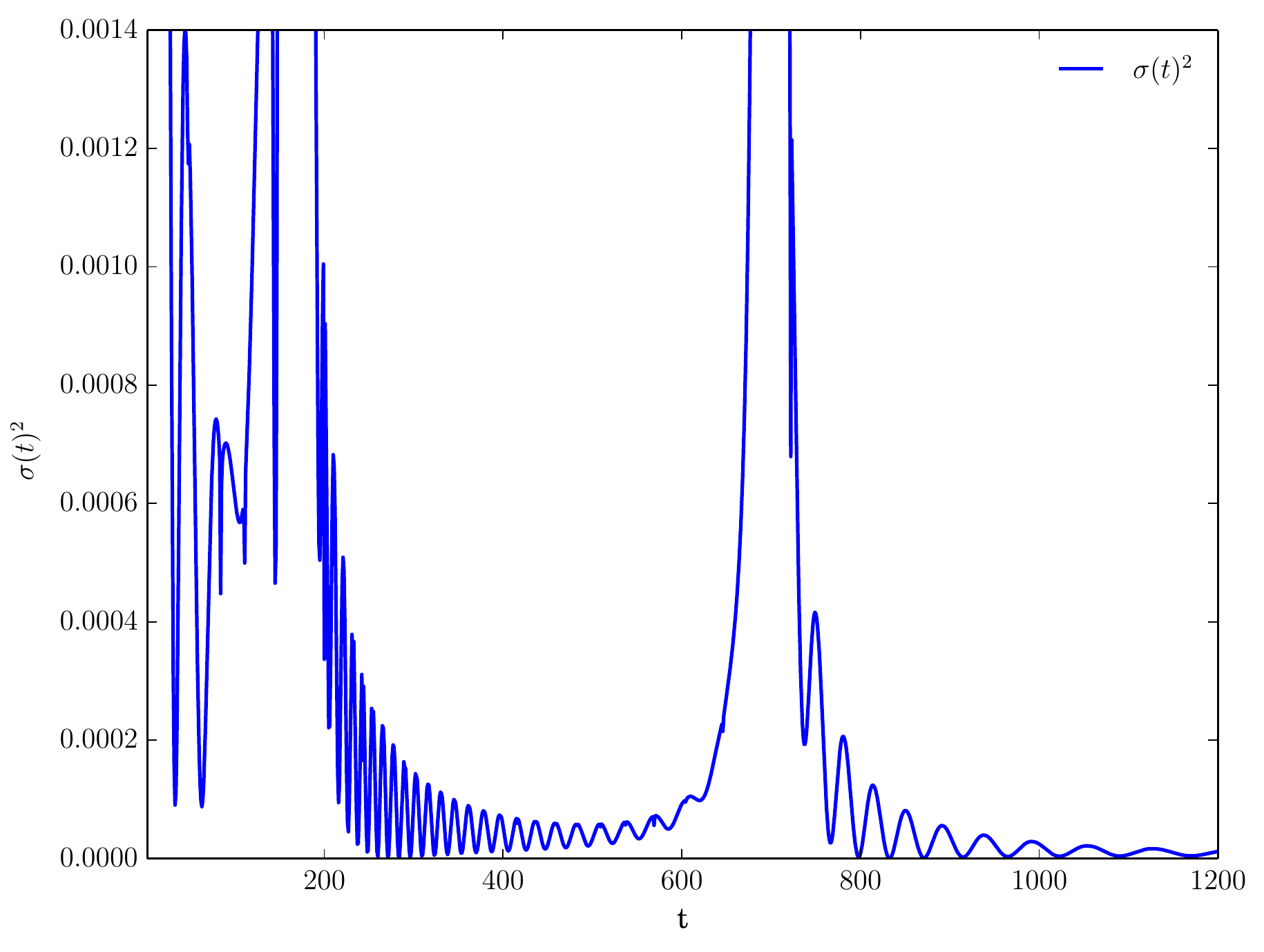}
\subcaption{\label{fig:entropyshear}}
\end{minipage}
\end{figure}

%%%%%%%%%%%%%%%%%%%%%%%

%%%%%%%%%%%%%%%%%%%%%%%%%%%%%%%%%%%%%%%%%%%%%%%%

\subsection{Evolution with non-comoving velocities}

To study the behaviour of this model under the influence of non-comoving
matter we assume that only the radiation field possesses non-comoving velocities
(i.e. the ghost field is comoving). In the case of Bianchi IX, we find that
the scale factors do undergo a bouncing behaviour, see Figure \ref%
{fig:entropyvolume}, as in the case without the non-comoving velocities. The
volume scale factor, $abc$, mimics the behaviour of cube of the scale factor
in the isotropic Friedmann case and shows an increase in height of its
expansion maxima as the entropy of the constituents is increased from cycle
to cycle. The individual scale factors oscillate out of phase with each
other and with different expansion maxima, similar to their behaviour
without the non-comoving velocities, see Figure \ref{fig:entropyindividual}.
However, the period of the volume oscillations is greater than in the
comoving velocities case. Thus, the model takes longer to recollapse on
average than in the comoving case, making each cycle last longer in comoving
proper $t$ time. \emph{\ }

%%%%%%shear and 3-curvature for entropy increase%%%%
%
%\begin{figure}[tbp]
%\caption{Evolution of (a) the $3$-curvature and (b)  the shear with the increase of entropy with time $t$ in a Bianchi IX universe where the radiation is not comoving with the
%tetrad frame, also containing a comoving dust field and a comoving ghost field to facilitate the
%bounce. }\centering\hfill \break 
%\begin{minipage}{0.48\textwidth}
%\includegraphics[width=1.0\linewidth, height=0.20\textheight]{curvature_entropy}
%\subcaption{\label{fig:entropycurvature}}
%\end{minipage}\hfill 
%\begin{minipage}{0.45\textwidth}
%\includegraphics[width=1.0\linewidth,height=0.20\textheight]{shear_entropy}
%\subcaption{\label{fig:entropyshear}}
%\end{minipage}
%\end{figure}
%
%%%%%%%%%%%%%%%%%%%%%%%%%%%%%%%%%%%%%%%%%%%%%%%%%

%%%%%velocities and sum of velocities for entropy increase%%%%

\begin{figure}[h]
\caption{(a), (b), and (c): Evolution of the squares of the 3-velocity components of non-comoving radiation with the increase in entropy in time $t$
in a Bianchi IX universe consisting of non-comoving radiation, as well as comoving
dust and the ghost fields, the latter to facilitate the bounce. Unlike in Figures \ref{fig:velconstraintu1},\ref{fig:velconstraintu2} and \ref{fig:velconstraintu3}, the velocity constraint equation \eqref{eq:vel_constraint} has not been explicitly imposed. The evolution of $u_2 (t)^2$ and $u_3(t)^2$ are highly oscillatory especially in the second cycle with very small time periods of oscillation, and to show this behaviour clearly, the plots are magnified and partly inset. }%
\centering\hfill \break 
\begin{minipage}{0.48\textwidth}
\includegraphics[width=1.1\linewidth, height=0.28\textheight]{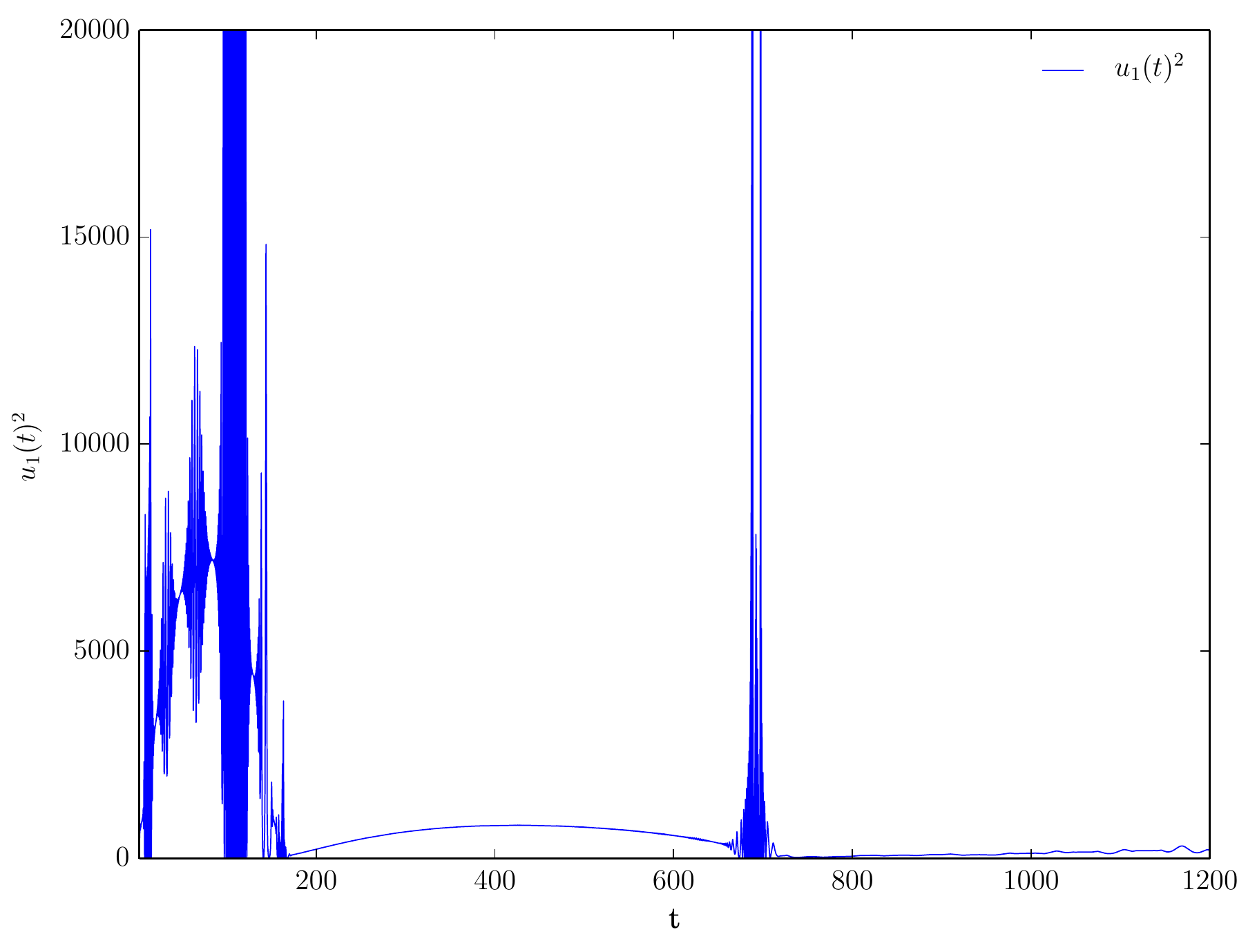}
\subcaption{\label{fig:velu1}}
\end{minipage}\hfill 
\newpage
\begin{minipage}{0.48\textwidth}
\includegraphics[width=1.0\linewidth,height=0.28\textheight]{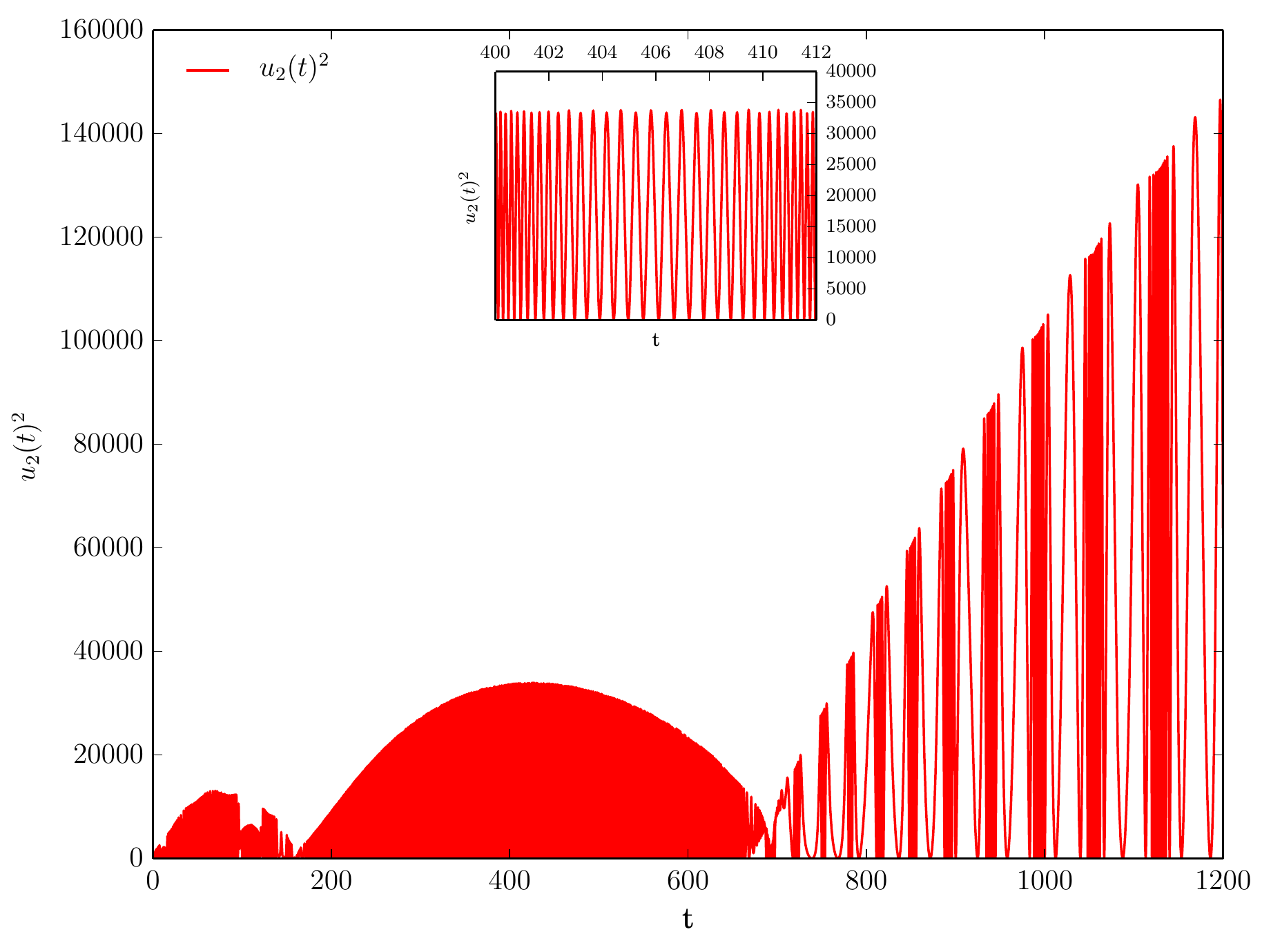}
\subcaption{\label{fig:velu2}}
\end{minipage}%
\begin{minipage}{0.45\textwidth}
\includegraphics[width=1.0\linewidth,height=0.28\textheight]{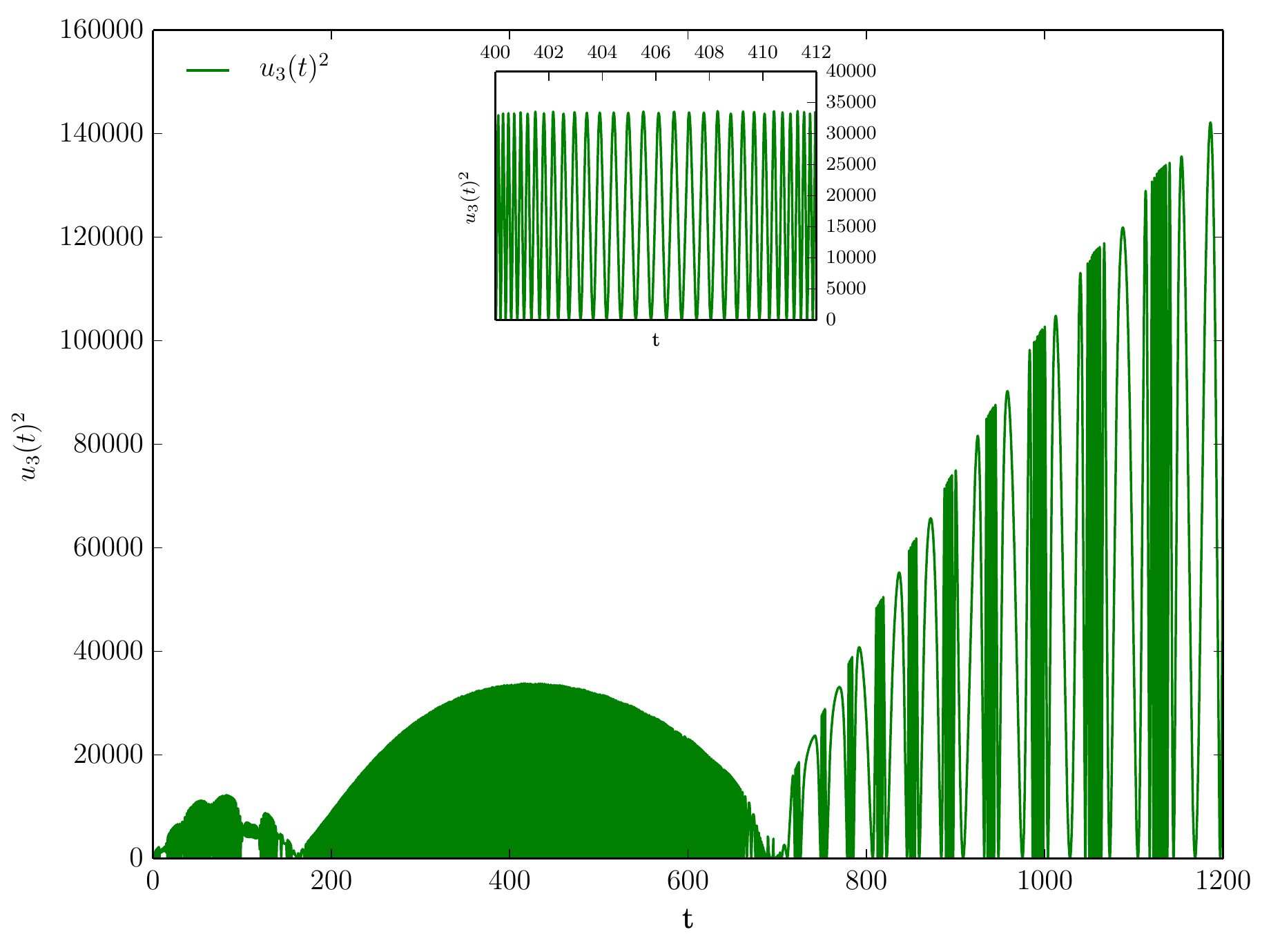}
\subcaption{\label{fig:velu3}}
\end{minipage}
\end{figure}

The shear and the $3$-curvature undergo oscillations which increase in
amplitude and frequency near the minima and do not appear to fall to smaller
and smaller values, see Figures \ref{fig:entropycurvature} and \ref%
{fig:entropyshear}.

The velocity components themselves show oscillatory behaviour, see Figures %
\ref{fig:velu1}, \ref{fig:velu2} and \ref{fig:velu3}. However, the amplitude
of their oscillations undergoes cyclic behaviour. The amplitudes of
oscillations fall to their smallest values at the expansion minima of the
scale factors. After the first oscillation, one of the velocity components
starts undergoing very small oscillations around a nearly constant value. We
give an approximate analytic analysis of this evolution in the Appendix. 

\pagebreak
\newpage
\section{The effects of a cosmological constant}

%%%%%%%%%%%%%%%%%%%%%%%%%%%%%%%%%%%%%%%%%%%%%%%%
%%%%%shear and 3-curvature for +ve cosmological constant%%%%

\begin{figure}[!ht]
\caption{Evolution of (a)  the shear, and (b)  the $3$-curvature (left to right) and
the individual scale factors with the increase in entropy with time  $t$ in a Bianchi IX universe where the radiation is not comoving with the
tetrad frame, containing a comoving dust field and a comoving ghost field to facilitate the
bounce, together with a positive cosmological constant. }\centering\hfill
\break 
\begin{minipage}{0.48\textwidth}
\includegraphics[width=1.1\linewidth, height=0.25\textheight]{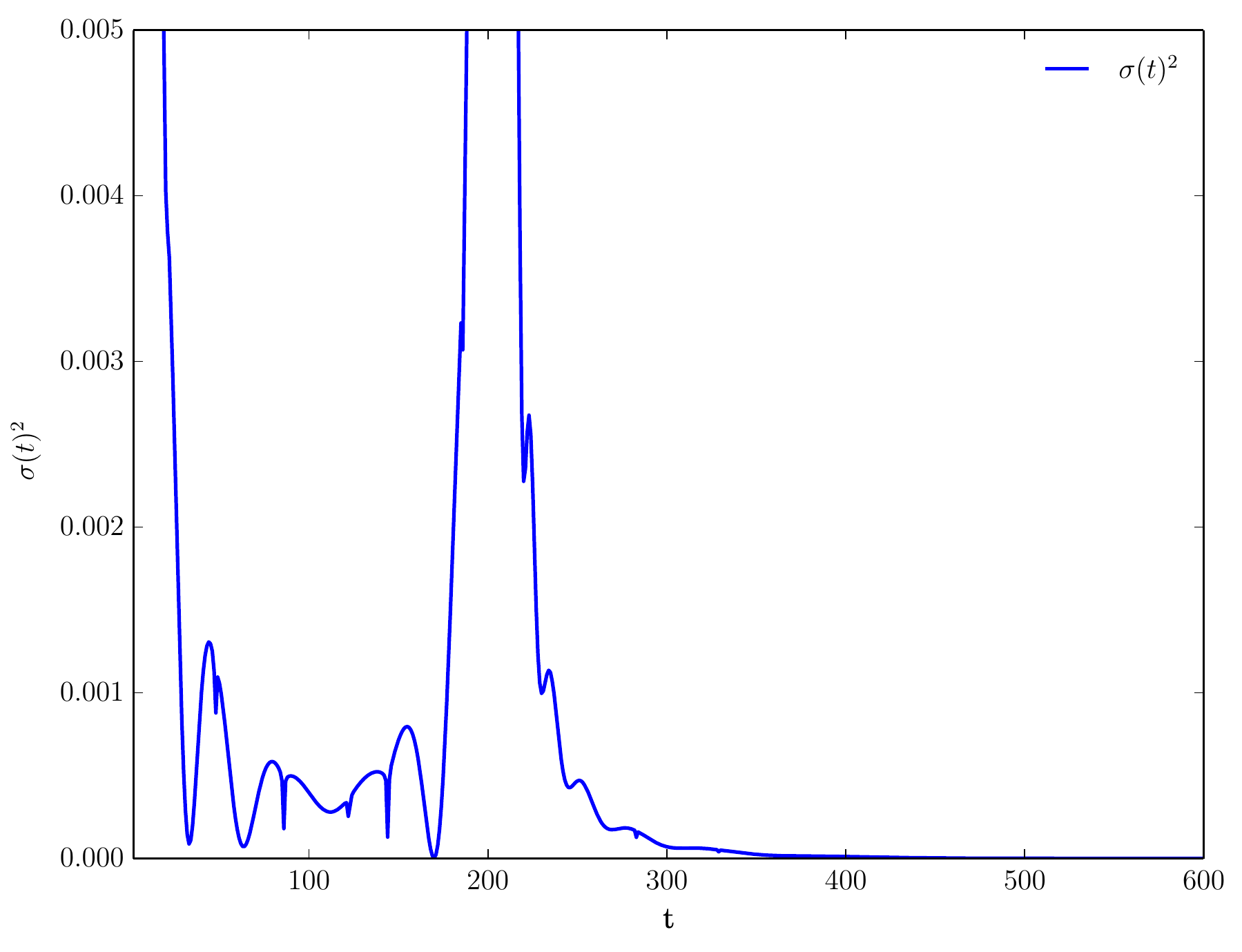}
\subcaption{\label{fig:lambdashear}}
\end{minipage}\hfill 
\begin{minipage}{0.45\textwidth}
\includegraphics[width=1.1\linewidth,height=0.25\textheight]{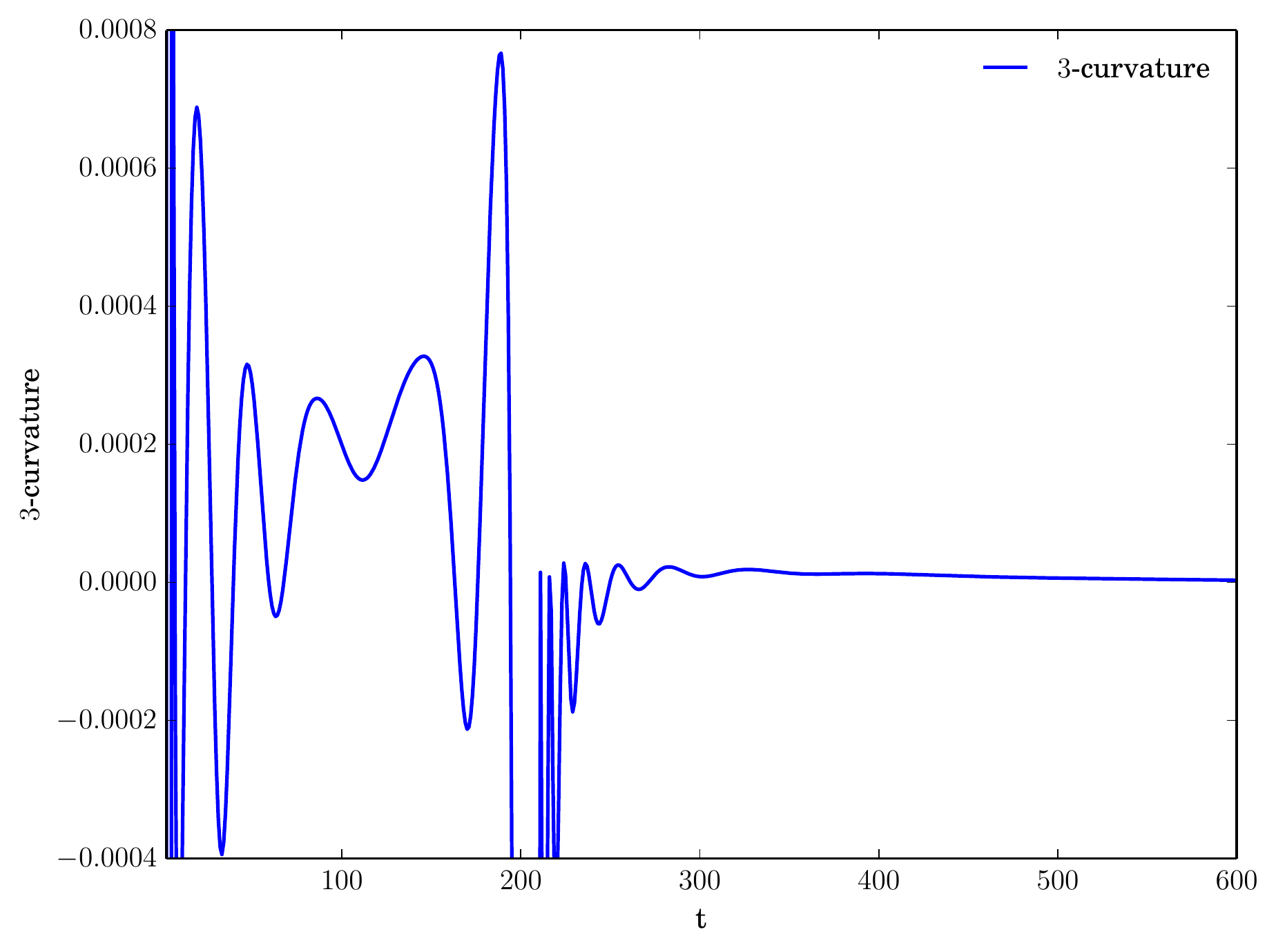}
\subcaption{\label{fig:lambdacurvature}}
\end{minipage}
\end{figure}

%%%%%%%%%%%%%%%%%%%%%%%%%%%%%%%%%%%%%%%%%%%%%%%%

\subsection{Positive cosmological constant ($\Lambda >0$)}

Now we add a cosmological constant to the model. The effect of cosmological
constant domination in the case of comoving velocities was to cause the
model to change from a cyclical behaviour to asymptotically de Sitter like
expansion \cite{me} (note that the cosmic no hair theorems \cite{nohair1,nohair2} do
not cover the type IX case because the 3-curvature scalar can be positive).

%%%%%individual scale factors and volume scale factors for +ve cosmological constant%%%%

\begin{figure}[tbp]
\caption{Evolution of (a) the volume scale factor, and (b) the individual directional Hubble
rates (left to right) with the increase in entropy, and a
positive cosmological constant, with time $t$ in a Bianchi IX universe where the
radiation is not comoving with the tetrad frame, also containing a comoving dust field and
a comoving ghost field to facilitate the bounce. The blue starred, red dotted, and
solid green lines correspond to derviatives of the principal values of the $3$-metric in the tetrad frame, Hubble rates $\dot{a}/a$, $\dot{b}/b$ and $\dot{c}/c$ respectively. The model undergoes approach to de Sitter expansion when the
cosmological constant eventually dominates the dynamics after cycles become large enough to ensure this.}%
\centering\hfill \break 
\begin{minipage}{0.48\textwidth}
\includegraphics[width=1.1\linewidth, height=0.25\textheight]{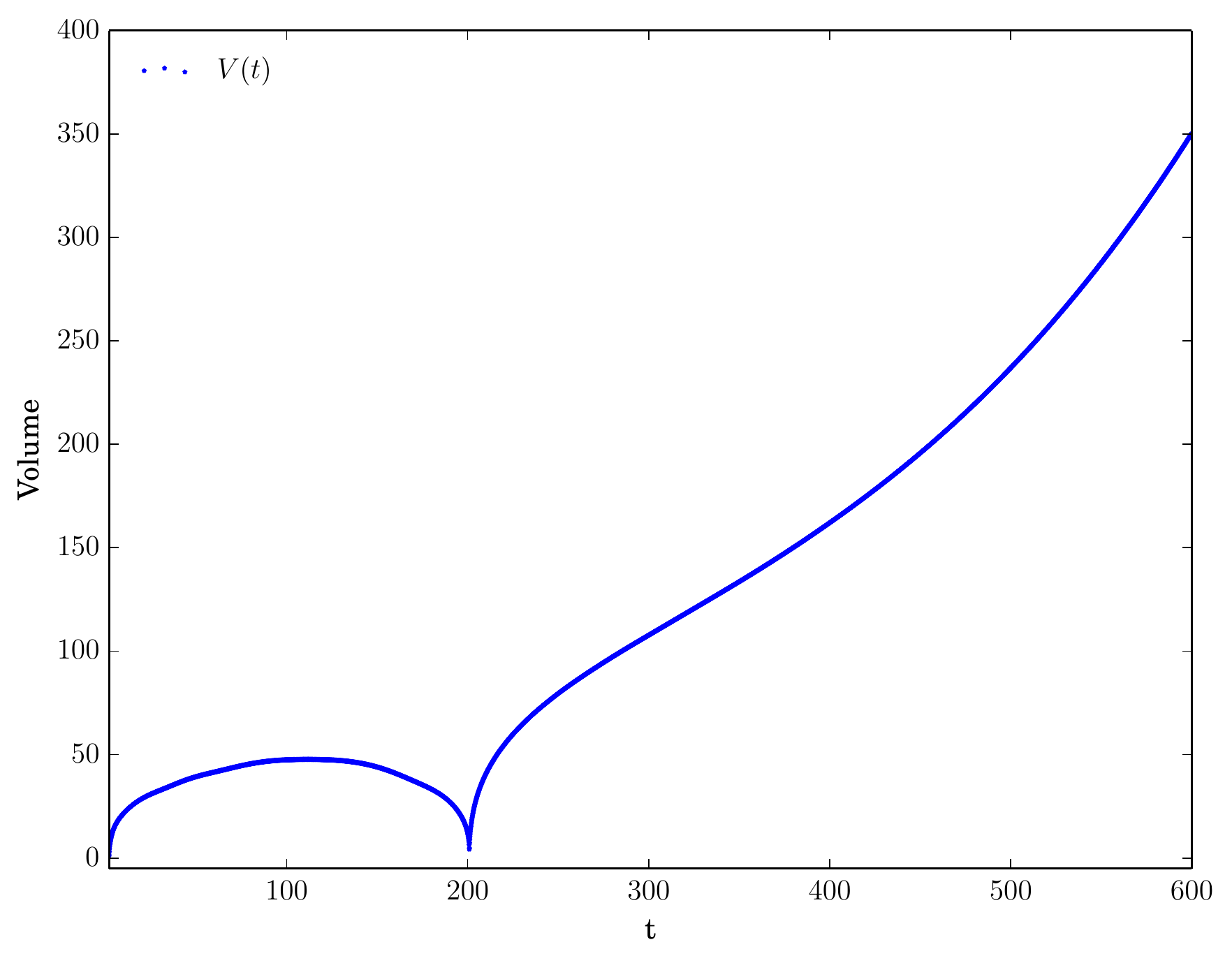}
\subcaption{\label{fig:lambdavolume}}
\end{minipage}\hfill 
\begin{minipage}{0.45\textwidth}
\includegraphics[width=1.1\linewidth,height=0.25\textheight]{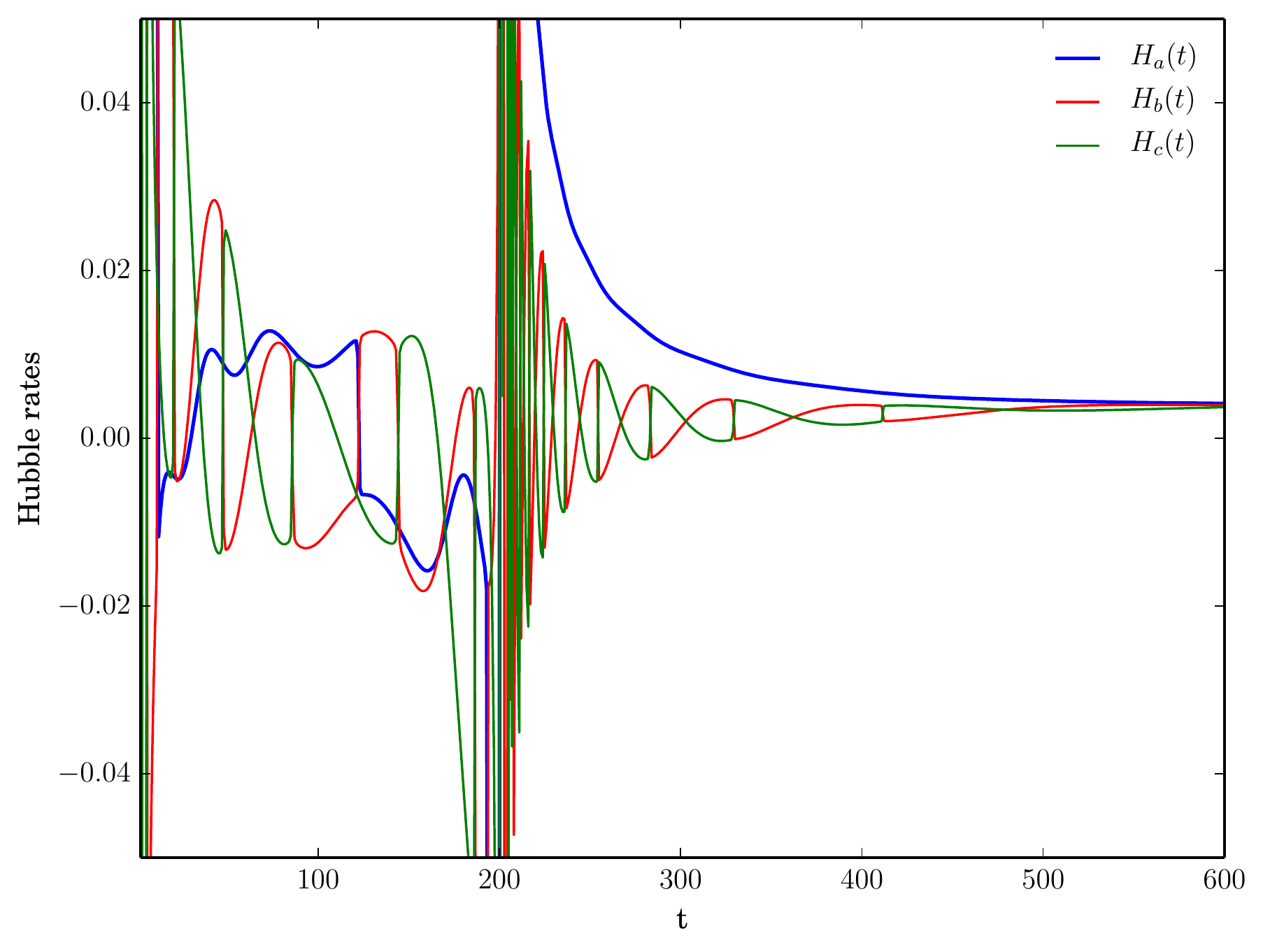}
\subcaption{\label{fig:lambdaindividual}}
\end{minipage}
\end{figure}

%%%%%%%%%%%%%%%%%%%%%%%%%%%%%%%%%%%%%%%%%%%%%%%%%
%%%%%%shear and 3-curvature for +ve cosmological constant%%%%
%
%\begin{figure}[!ht]
%\caption{Evolution of (a)  the shear, and (b)  the $3$-curvature (left to right) and
%the individual scale factors with the increase in entropy withtime  $t$ in a Bianchi IX universe where the radiation is not comoving with the
%tetrad frame, containing a comoving dust field and a comoving ghost field to facilitate the
%bounce, together with a positive cosmological constant. }\centering\hfill
%\break 
%\begin{minipage}{0.48\textwidth}
%\includegraphics[width=1.1\linewidth, height=0.25\textheight]{shear_lambda}
%\subcaption{\label{fig:lambdashear}}
%\end{minipage}\hfill 
%\begin{minipage}{0.45\textwidth}
%\includegraphics[width=1.1\linewidth,height=0.25\textheight]{curvature_lambda}
%\subcaption{\label{fig:lambdacurvature}}
%\end{minipage}
%\end{figure}
%
%%%%%%%%%%%%%%%%%%%%%%%%%%%%%%%%%%%%%%%%%%%%%%%%%
As in case with comoving velocities, the model is able to undergo
cyclical behaviour until the maxima grow large enough for the cosmological
constant to dominate at late times and then the dynamics approach a phase of
quasi de Sitter expansion, see Figure \ref{fig:lambdavolume}. The individual
expansion rates oscillate while the model is still undergoing cyclical
behaviour but approach a constant value $H_{0}=\sqrt{\Lambda /3}$ signalling
the onset of isotropic de Sitter behaviour, see Figure \ref%
{fig:lambdaindividual}.

In the de Sitter phase the shear and the curvature are diluted by expansion,
as expected, and fall exponentially rapidly to very small values, see
Figures \ref{fig:lambdashear} and \ref{fig:lambdacurvature}. The 3-curvature
can be seen to change sign from negative values (when the dynamics are far
from isotropy) to positive values (when the dynamics are close to isotropy).
Positive 3-curvature is necessary for a volume maximum to occur.

%%%%%velocities and sum of velocities for cosmological constant%%%%

\begin{figure}[tbp]
\caption{Evolution of the squares of velocity components of  non-comoving radiation  with the increase in entropy with time $t$
in a Bianchi IX universe consisting of non-comoving radiation, as well as a comoving
dust field and a comoving ghost field to facilitate the bounce, and a positive
cosmological constant. The graphs (a), (b) and (c) plot  the squares of the spatial components of the $4$-velocity in
the tetrad frame, $u_1 (t)^2$,$u_2 (t)^2$, and $u_3(t)^2$, respectively. }%
\centering\hfill \break 
\begin{minipage}{0.48\textwidth}
\includegraphics[width=1.1\linewidth, height=0.28\textheight]{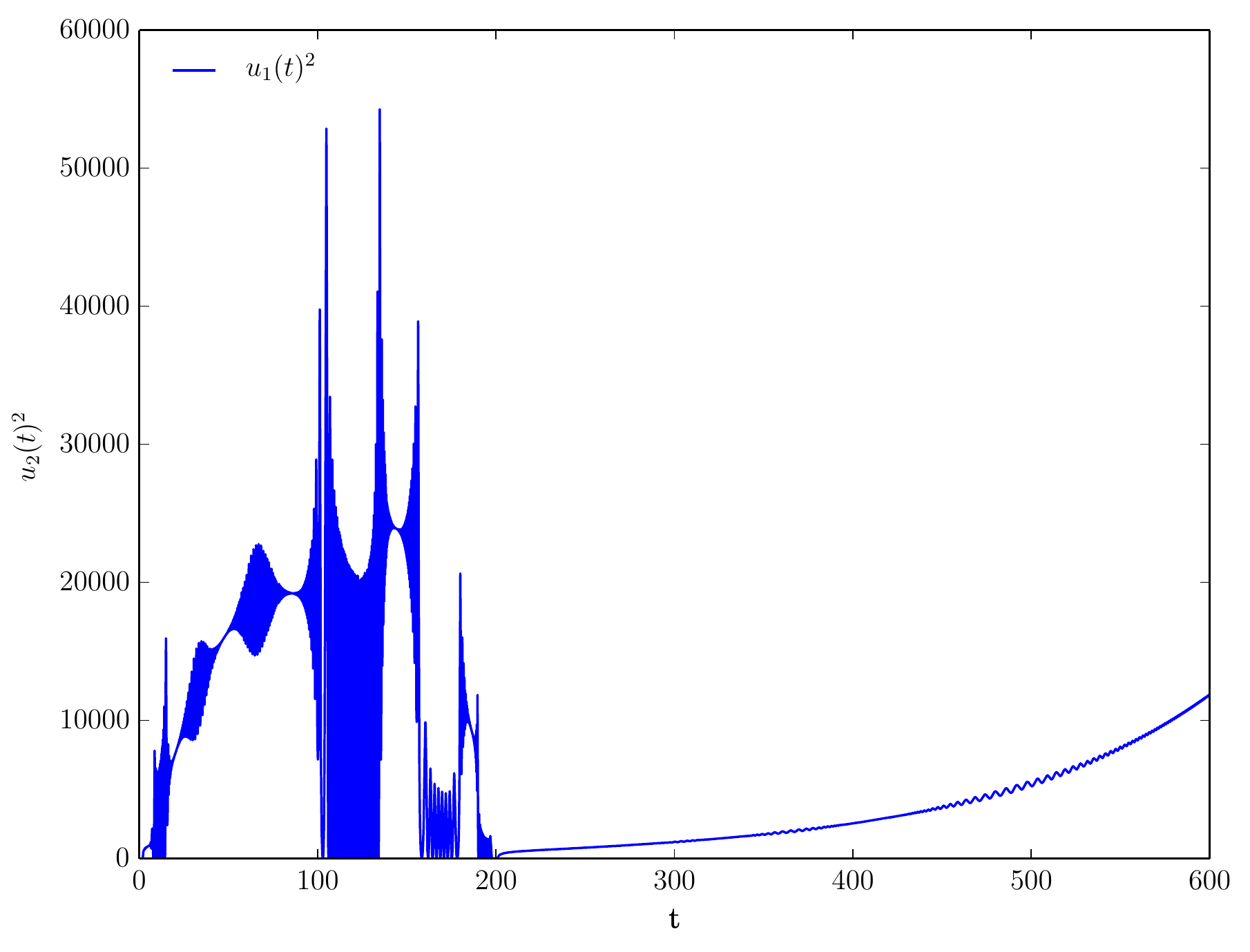}
\subcaption{\label{fig:lambdavelu1}}
\end{minipage}\hfill 
\begin{minipage}{0.48\textwidth}
\includegraphics[width=1.1\linewidth, height=0.28\textheight]{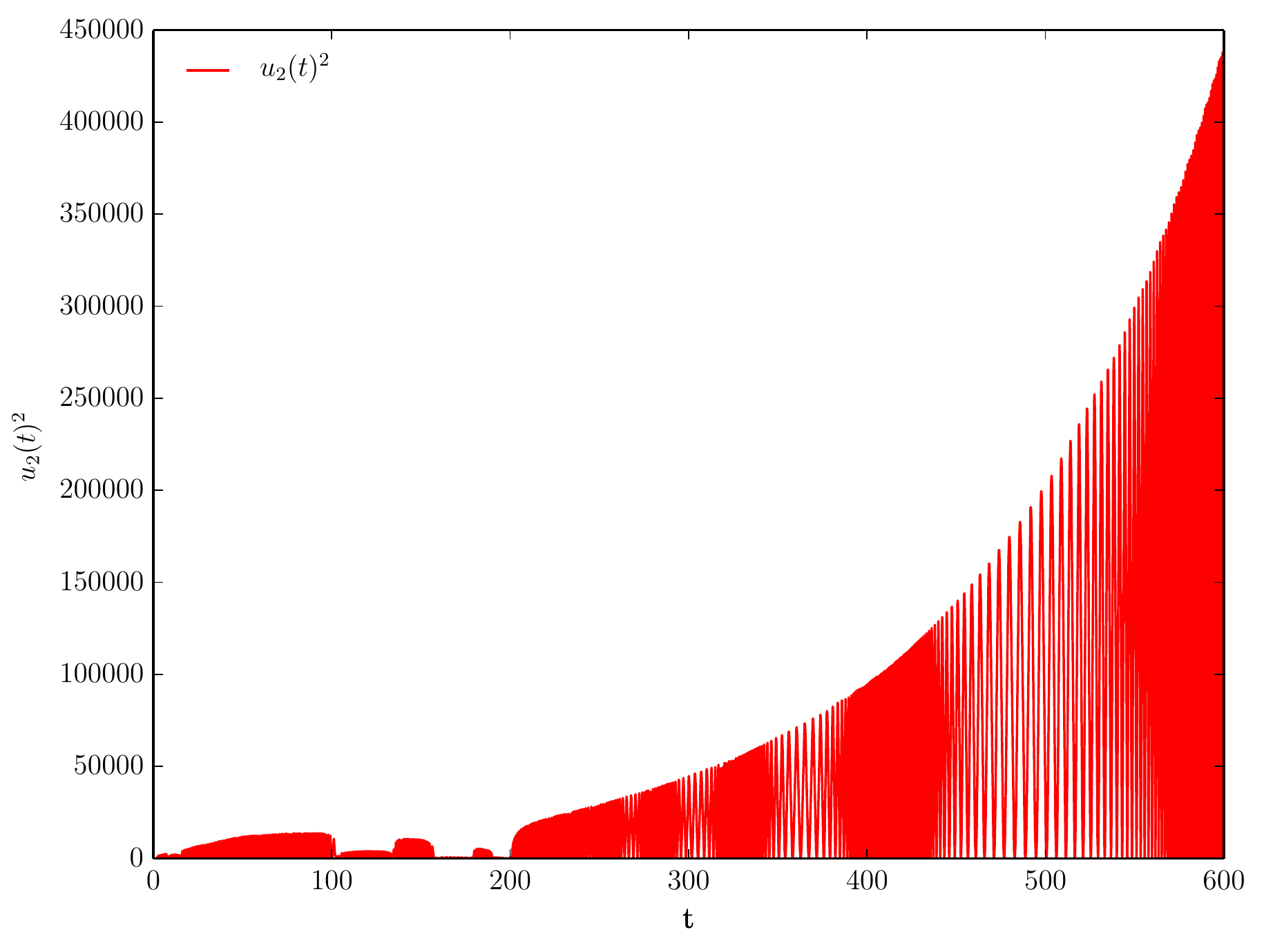}
\subcaption{\label{fig:lambdavelu2}}
\end{minipage}\hfill 
\begin{minipage}{0.48\textwidth}
\includegraphics[width=1.1\linewidth, height=0.28\textheight]{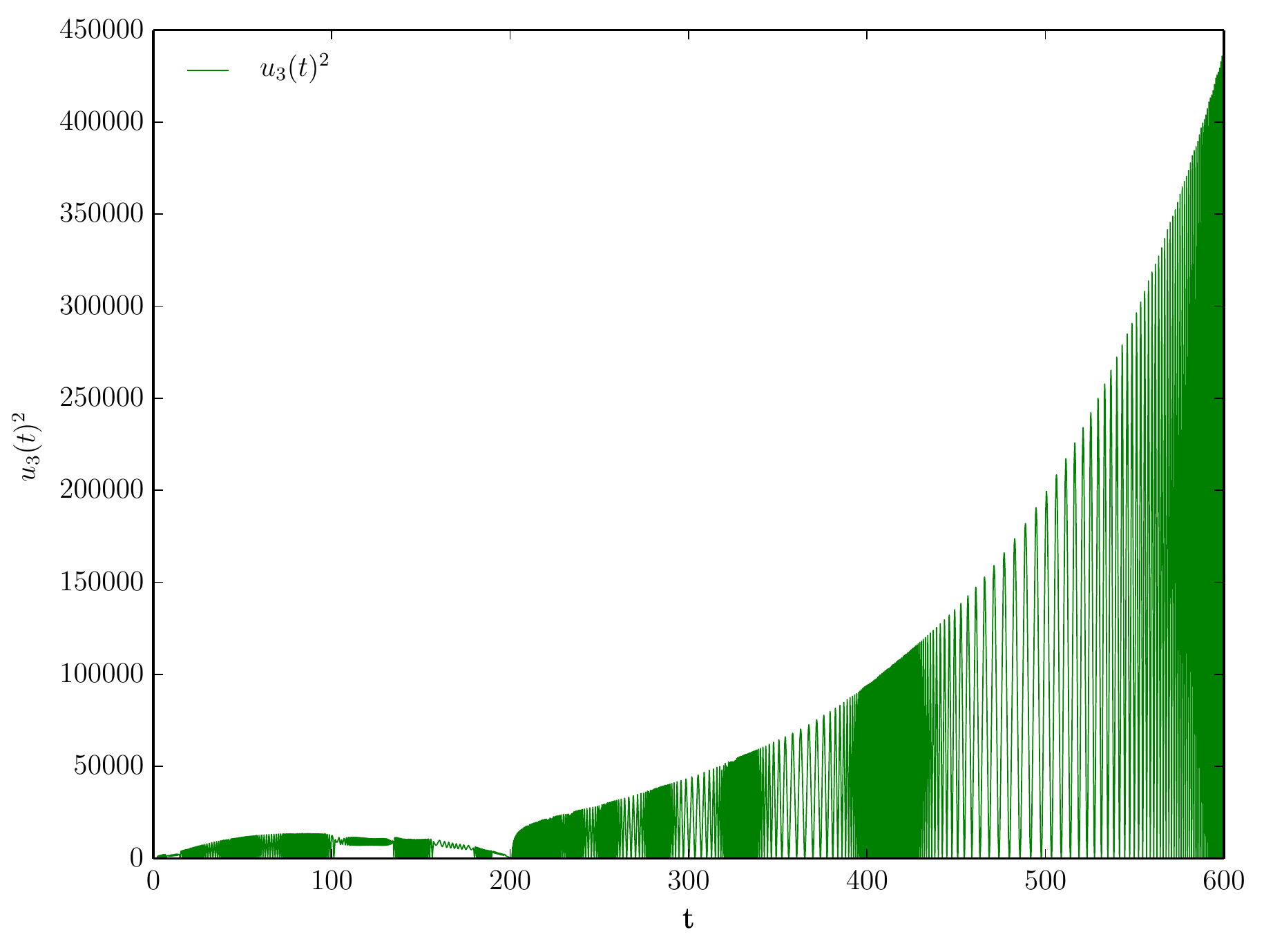}
\subcaption{\label{fig:lambdavelu3}}
\end{minipage}
\end{figure}

%%%%%%%%%%%%%%%%%%%%%%%%%%%%%%%%%%%%%%%%%%%%%%%%

The velocities themselves oscillate rapidly in each cycle, while the
amplitudes of the oscillations rise or fall according to the growth or
regression of the scale factors, see Figures \ref{fig:lambdavelu1}, \ref%
{fig:lambdavelu2} and \ref{fig:lambdavelu3}. The amplitudes of the
oscillations of the velocities in two of the directions grow with the
exponential expansion of the scale factors. As the scale factors expand
further, the time period of the oscillations of the velocities also
increases.

\subsection{Negative cosmological constant ($\Lambda <0$)}

%%%%%%%%%%%%%%%%%%%%%%%%%%%%%%%%%%%%%%%%%%%%%%%%
%%%%%shear and 3-curvature for -ve cosmological constant%%%%

\begin{figure}[!ht]
\caption{Evolution of (a) the shear, and (b)  the $3$-curvature with the increase of entropy with time $t$ in a Bianchi IX universe where the radiation is not comoving with the
tetrad frame, as well as comoving dust and ghost field, the latter to facilitate the
bounce, in the presence of a negative cosmological constant. }\centering%
\hfill \break 
\begin{minipage}{0.48\textwidth}
\includegraphics[width=1.1\linewidth, height=0.28\textheight]{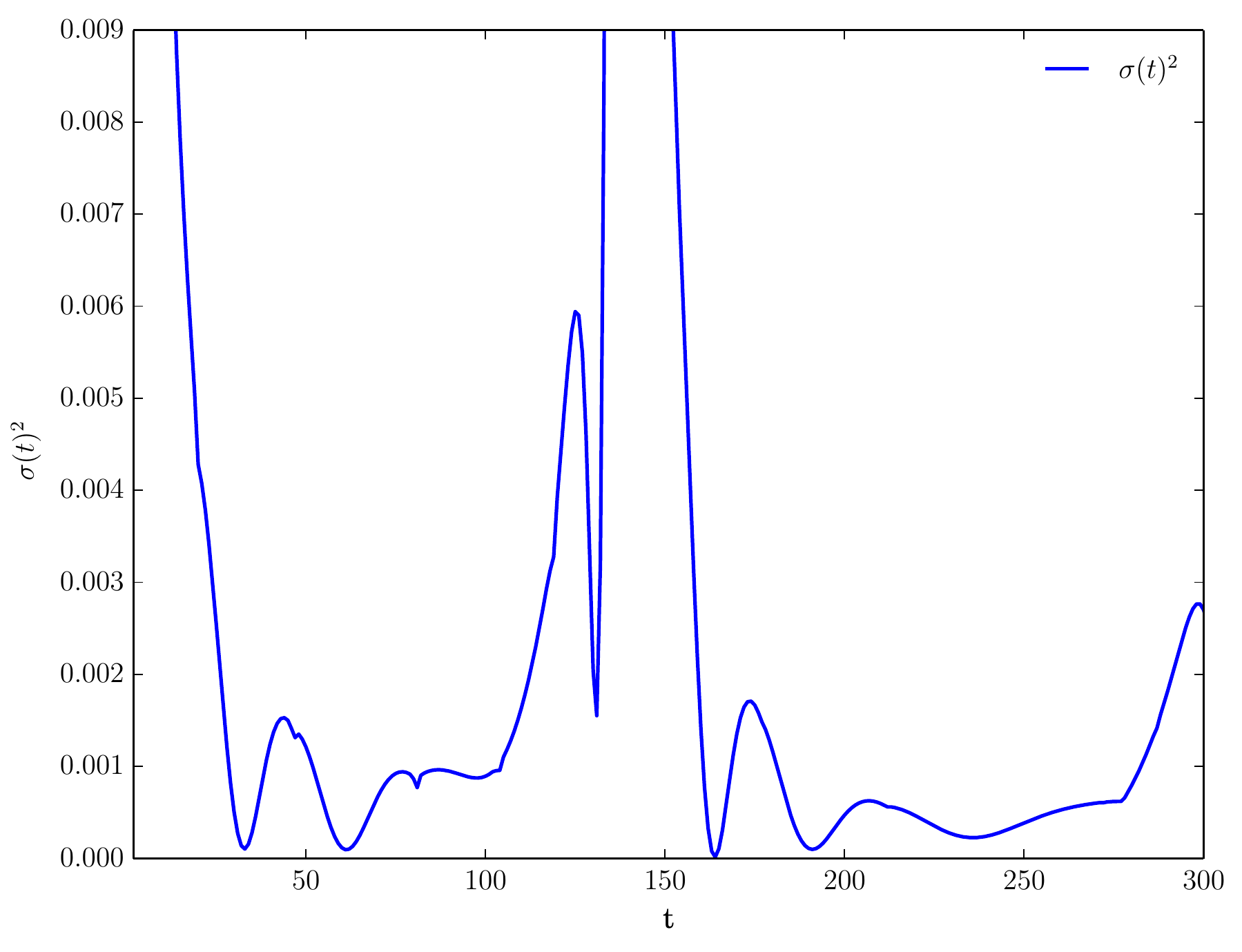}
\subcaption{\label{fig:SHUshear}}
\end{minipage}\hfill 
\begin{minipage}{0.45\textwidth}
\includegraphics[width=1.1\linewidth,height=0.28\textheight]{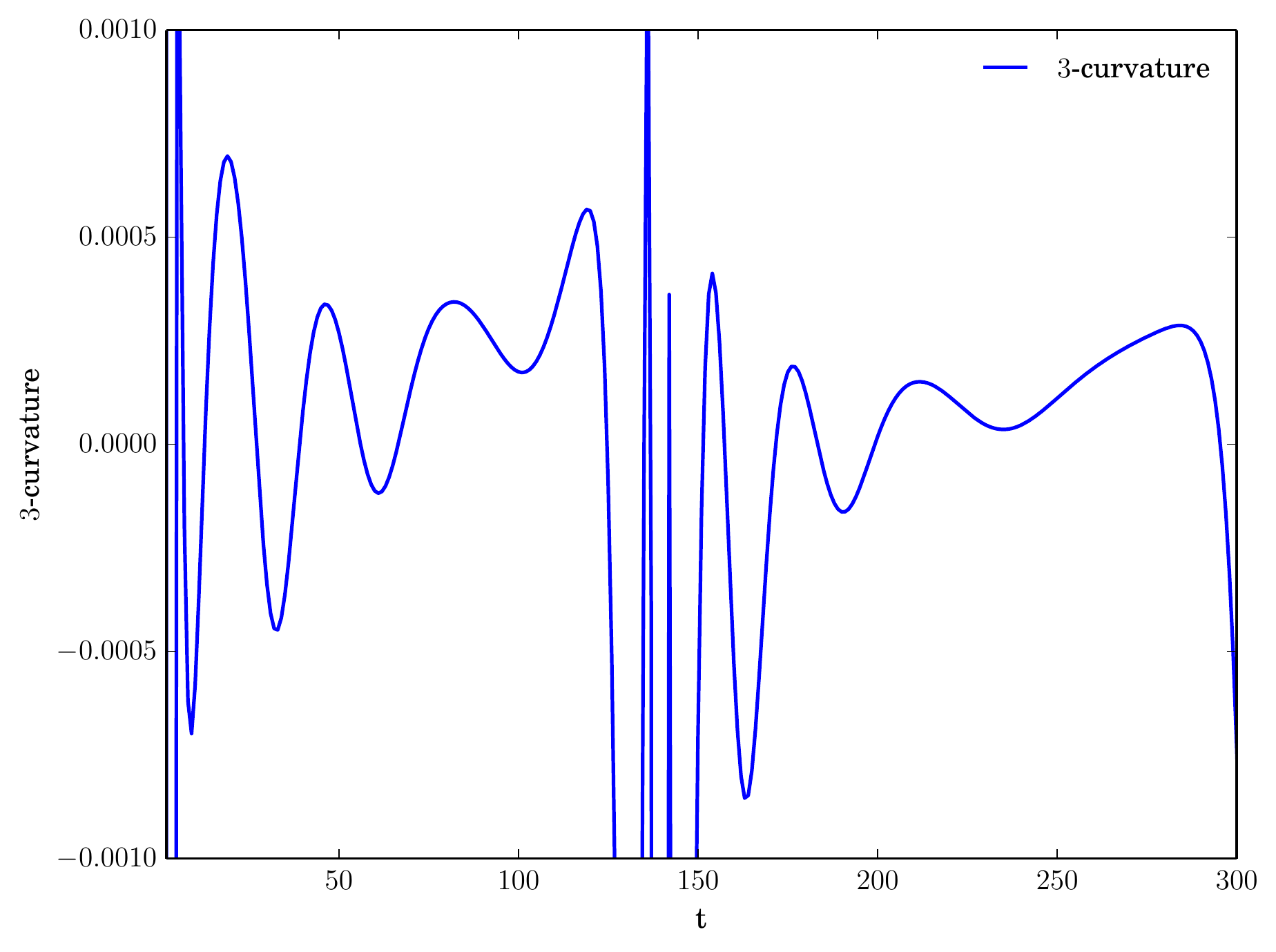}
\subcaption{\label{fig:SHUcurvature}}
\end{minipage}
\end{figure}
\begin{figure}[tbp]
\caption{Evolution of (a) the volume scale factor and (b) the individual scale
factors with $t$ in the presence of a negative cosmological constant  in a
Bianchi IX universe where the radiation is not comoving with the tetrad
frame, and containing a comoving  dust field and a comoving ghost field to facilitate the bounce.
The blue starred, red dotted and green solid lines correspond to the
principal values of the $3$-metric in the tetrad frame, scale factors $a(t)$, $b(t)$ and $c(t)$, respectively. }%
\centering\hfill \break 
\begin{minipage}{0.48\textwidth}
\includegraphics[width=1.1\linewidth, height=0.28\textheight]{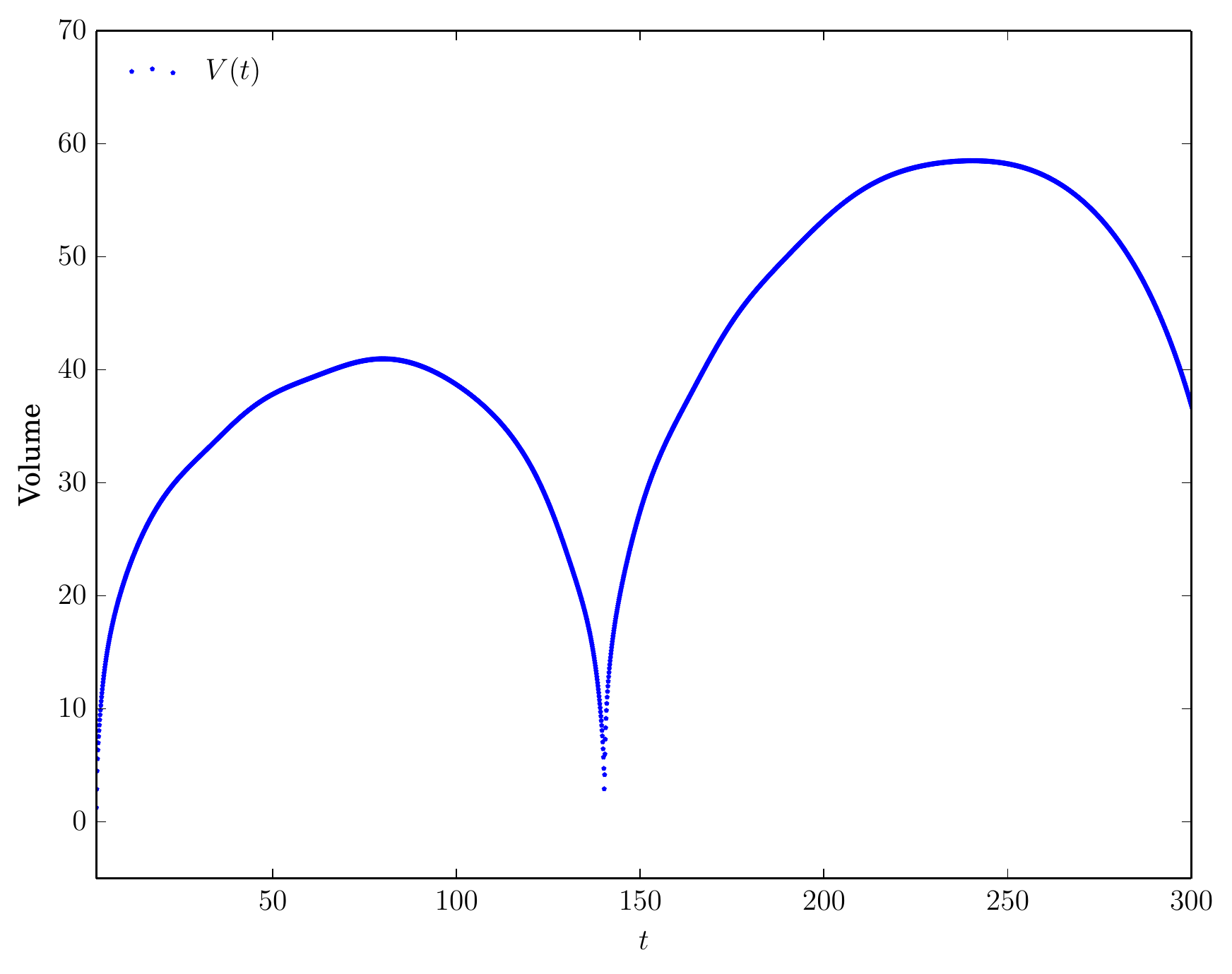}
\subcaption{\label{fig:SHUvolume}}
\end{minipage}\hfill 
\begin{minipage}{0.45\textwidth}
\includegraphics[width=1.1\linewidth,height=0.28\textheight]{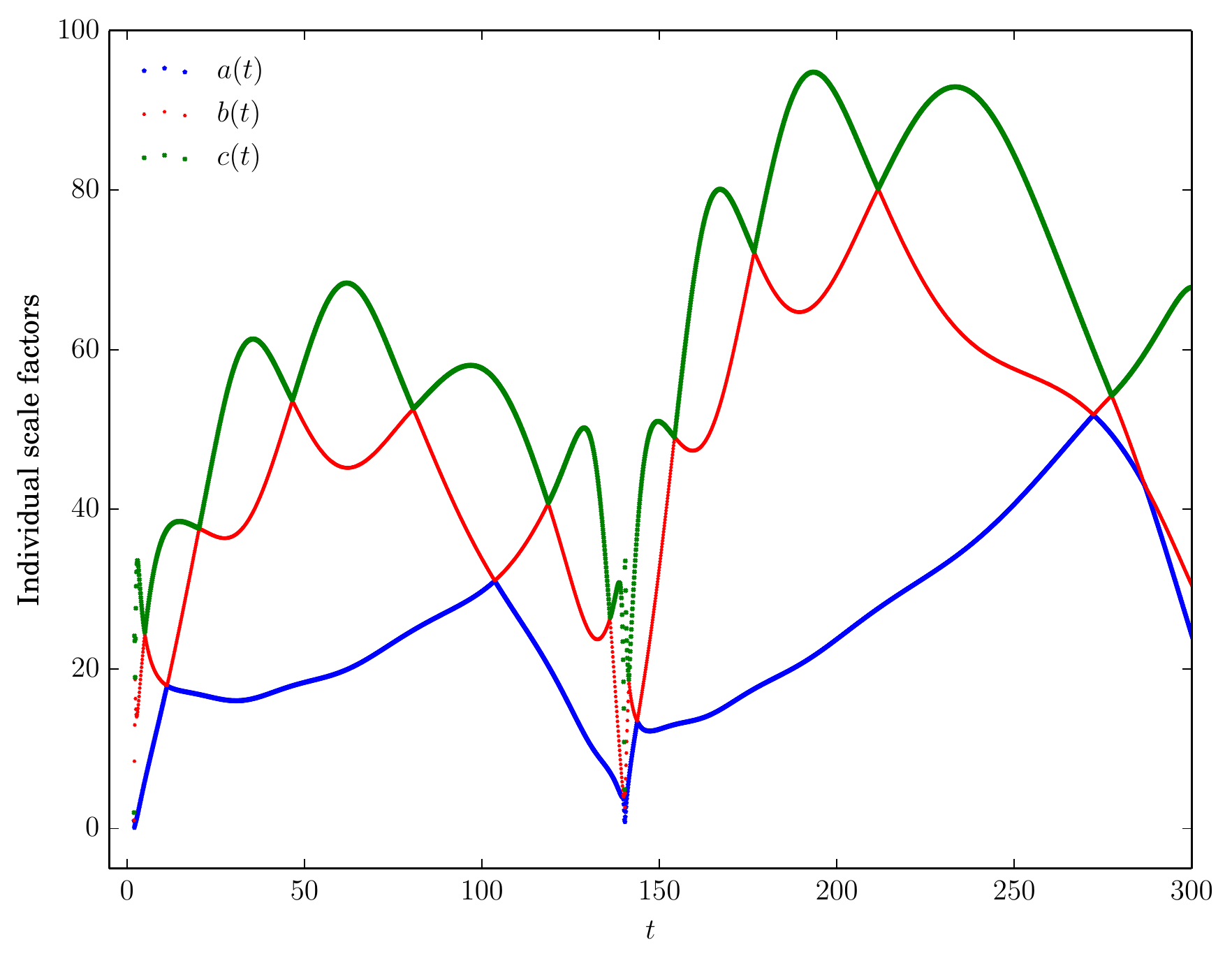}
\subcaption{\label{fig:SHUindividual}}
\end{minipage}
\end{figure}

%%%%%%%%%%%%%%%%%%%%%%%%%%%%%%%%%%%%%%%%%%%%%%%%%%
Adding a negative cosmological constant results in the universe always
recollapsing \cite{tip}, as this is just another null energy condition
violating field. For the behaviour of the volume and individual scale
factors, see Figures \ref{fig:SHUvolume} and \ref{fig:SHUindividual}.

%%%%%individual scale factors and volume scale factors for -ve cosmological constant%%%%

%%%%%%%%%%%%%%%%%%%%%%%%%%%%%%%%%%%%%%%%%%%%%%%%

%%%%%velocities and sum of velocities for cosmological constant%%%%

\begin{figure}[tbp]
\caption{Evolution of the squares of the velocities of non-comoving radiation with time $t$ in a Bianchi IX universe consisting of
non-comoving radiation, as well as comoving dust and ghost fields, the latter to
facilitate the bounce, and a negative cosmological constant. Plots (a), (b) and (c) show the squares of the spatial
components of the $4$-velocity in the tetrad frame, $u_1 (t)^2$,$u_2 (t)^2$, and  $u_3(t)^2$, respectively. The highly oscillatory behaviour of the velocity components with very short time period is captured by the magnified insets in each of the plots. }%
\centering\hfill \break 
\begin{minipage}{0.48\textwidth}
\includegraphics[width=1.1\linewidth, height=0.28\textheight]{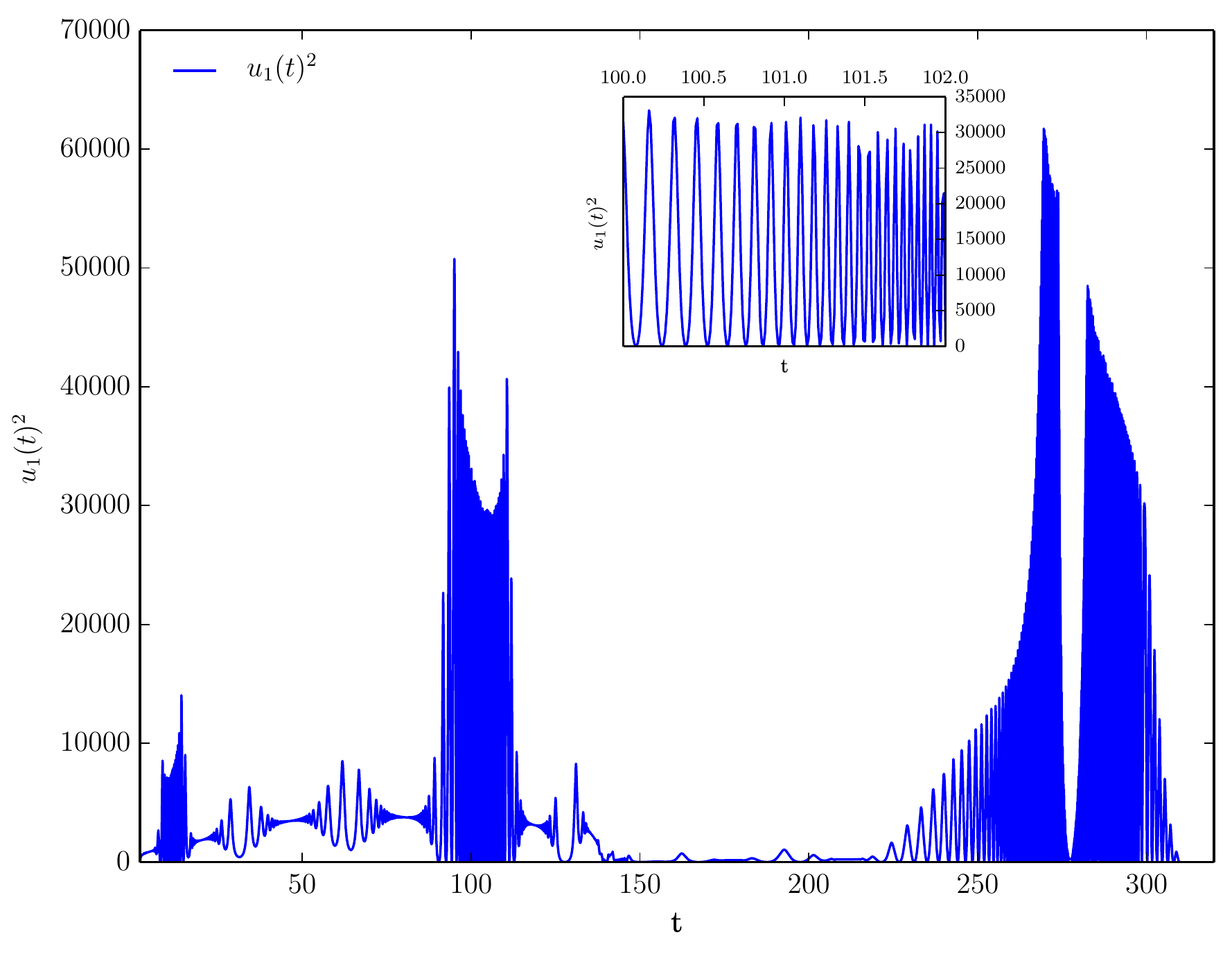}
\subcaption{\label{fig:SHUvelu1}}
\end{minipage}\hfill 
\begin{minipage}{0.48\textwidth}
\includegraphics[width=1.1\linewidth, height=0.28\textheight]{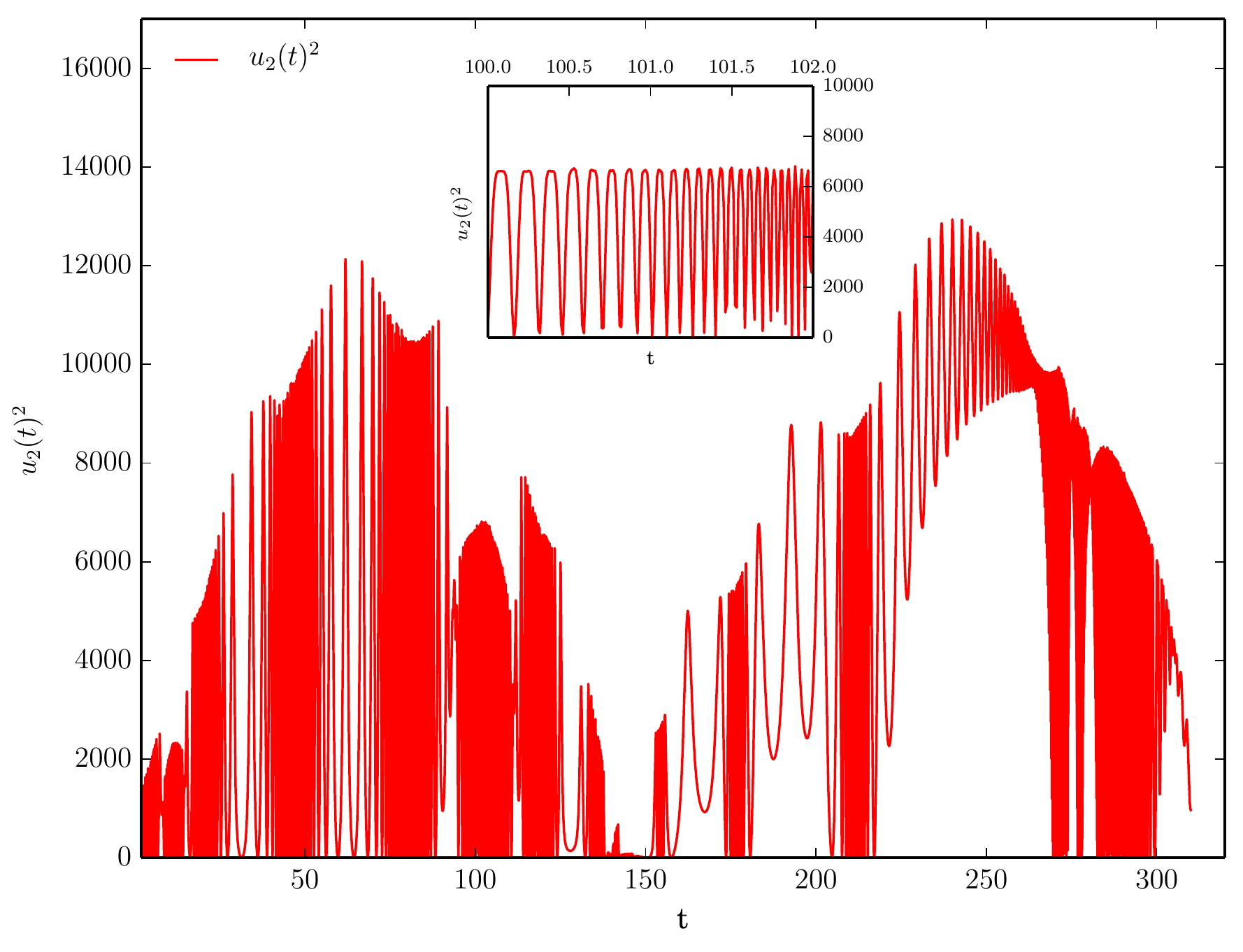}
\subcaption{\label{fig:SHUvelu2}}
\end{minipage}\hfill 
\begin{minipage}{0.48\textwidth}
\includegraphics[width=1.1\linewidth, height=0.28\textheight]{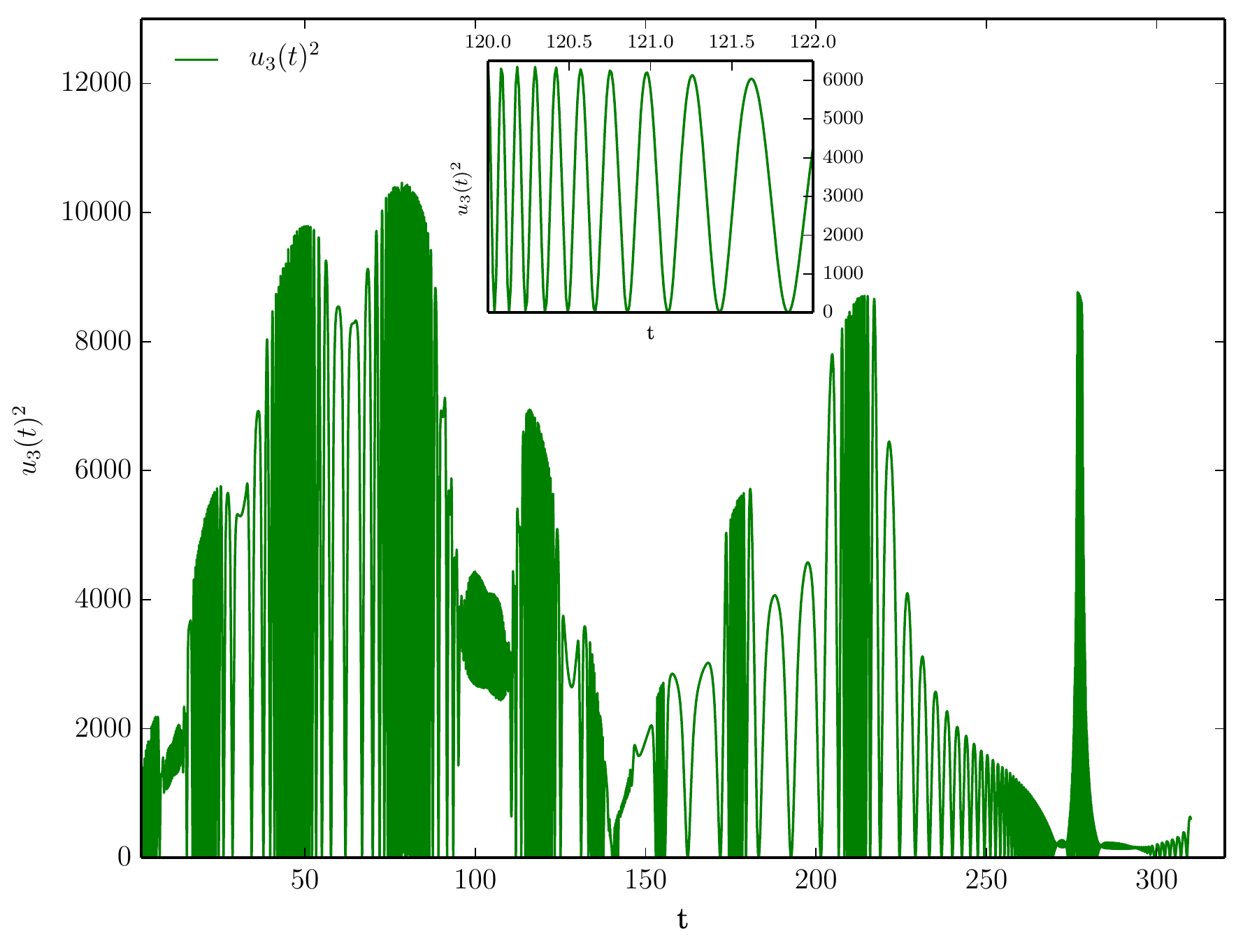}
\subcaption{\label{fig:SHUvelu3}}
\end{minipage}\hfill 
\end{figure}

%%%%%%%%%%%%%%%%%%%%%%%%%%%%%%%%%%%%%%%%%%%%%%%%

%%%%%%shear and 3-curvature for -ve cosmological constant%%%%
%
%\begin{figure}[!ht]
%\caption{Evolution of (a) the shear, and (b)  the $3$-curvature with the increase of entropy with time $t$ in a Bianchi IX universe where the radiation is not comoving with the
%tetrad frame, as well as comoving dust and ghost field, the latter to facilitate the
%bounce, in the presence of a negative cosmological constant. }\centering%
%\hfill \break 
%\begin{minipage}{0.48\textwidth}
%\includegraphics[width=1.1\linewidth, height=0.28\textheight]{shear_SHU}
%\subcaption{\label{fig:SHUshear}}
%\end{minipage}\hfill 
%\begin{minipage}{0.45\textwidth}
%\includegraphics[width=1.1\linewidth,height=0.28\textheight]{curvature_SHU}
%\subcaption{\label{fig:SHUcurvature}}
%\end{minipage}
%\end{figure}

The ghost field allows the model to undergo more cycles of oscillation. As
we are not introducing an increase in entropy and all the cycles are of
equal size, we shall focus on one cycle. The velocities all oscillate and
increase with the volume of the universe. One of the velocities ($%
u_{1}(t)^{2}$) also oscillates but with smaller amplitude around a constant
value. Only at the end of each cycle does this velocity component show an
increase in the amplitude of oscillations, see Figures \ref{fig:SHUvelu1}, %
\ref{fig:SHUvelu2} and \ref{fig:SHUvelu3}.

%%%%%%%%%%%%%%%%%%%%%%%%%%%%%%%%%%%%%%%%%%%%%%%%

The shear and the $3$-curvature undergo oscillations, falling to their
smallest values at the moments when the volume of the universe is at its
highest, see Figures \ref{fig:SHUshear} and \ref{fig:SHUcurvature}. Again,
we see the 3-curvature taking on negative values when the dynamics are
significantly anisotropic and positive values when close to isotropy.

\section{Conclusions}

To complete the analysis of the shape of cyclic closed anisotropic
universes, it is important to include the effects of non-comoving matter. In
the current analysis, we have extended the results of \cite{me} by including
a radiation field that is not comoving with the reference tetrad frame. This
tilted velocity field introduces vorticity, in addition to the shear and $3$%
-curvature anisotropies, into the universe.

We found that, as in the comoving case, the expansion maxima increases with
increasing entropy of the constituents from cycle to cycle, while the
individual scale factors oscillate out of phase with each other. The overall
dynamics approach flatness over many cycles but they become increasingly
anisotropic. We find a new effect in oscillating universes with non-comoving
velocities and vorticity. Over successive cycles of entropy increase the
conservation of momentum and angular momentum ensures that there is a
decrease in the magnitude of the velocities and vorticities in response to
the increase of entropy. We modelled entropy increase per cycle by adding
entropy at the start of each cycle of a closed universe. We also included a
comoving ghost field with negative energy density in order to create a
bounce at finite expansion minima and avoid the chaotic mixmaster regime as $%
t\rightarrow 0$ -- it is not relevant in practice to post-Planck time
evolution.

Our numerical study shows that the velocities oscillate many times and
around an almost constant value per cycle, and the amplitude of the
oscillations increases with the increase in expansion volume. The velocity
in one of the directions  tends to a constant value after initially
undergoing several oscillations. On explicitly imposing the constraint
equation arising out of particle number conservation, see equation %
\eqref{eq:vel_constraint}, we find that an increase in entropy density(and
hence energy density as for radiation $s\propto \rho ^{3/4}$) produces a
corresponding decrease in the components of the non-comoving velocity, and
vice versa. 

When we add a positive cosmological constant to a model containing radiation
and a ghost field we confirm that the oscillations are sustained until the
cosmological constant dominates the dynamics, after which the scale factors
enter a period of quasi de Sitter expansion. The velocities oscillate with
amplitude increasing with increasing scale factor as before, but after
cosmological constant domination, the time period of oscillations starts
increasing, and they oscillate less rapidly, around a constant value. The
asymptotic state is de Sitter with a constant velocity field.

When we add a negative cosmological constant we find there is always
collapse, as expected. Studying one cycle we see that the scale factors
oscillate out of phase with each other. The velocities in two directions
oscillate with increasing amplitude as the volume increases but decrease
again with decreasing volume. The velocity in the third direction, however,
oscillates with very small amplitude around a constant value, only
increasing in oscillation amplitude at the end of each cycle when the volume
is its smallest.

We conclude that the inclusion of non-comoving velocities has the effect of
increasing the time period of the oscillations of the model. The velocities
oscillate rapidly per cycle but with increasing amplitude as the volume of
the universe increases, in at least two directions. In the third direction,
the velocity oscillates around a constant value with very small amplitude,
and hence remains nearly constant per cycle. It only increases in amplitude
when the model collapses, before relapsing again to a nearly constant value
during the next cycle. Our analysis has identified the principal ingredients
of a general cyclic closed universe in the case of spatial homogeneity. In a
future work we will explore the effects of inhomogeneities on these
conclusions.

\newpage \appendix

\section{Approximate analysis of the radiation era}

We seek an approximate solution of the velocity evolution equations in the
type IX model during the radiation era. In our earlier study \cite{me}
without velocities we found a long period of evolution during the radiation
era (before the curvature creates slow-down of the expansion near the volume
maximum) with the scale factors evolving to a good approximation in a
quasi-axisymmetric manner during the radiation era, as

\begin{equation}
a(t)=a_{0}t^{1/2}[\ln (t)]^{-1/2},b(t)=b_{0}t^{1/2}[\ln
(t)]^{-1/2},c(t)=c_{0}t^{1/2}\ln (t).  \label{scale}
\end{equation}

When the effects of the velocities in the Bianchi type IX radiation universe
are small they can be treated as test motions on an expanding radiation
background governed by equations \eqref{eq:current_conservation} and %
\eqref{eq:v0_constraint}. We examine a typical case where we choose

\begin{equation*}
v_{3}=\mathrm{constant.}
\end{equation*}%
This is consistent with the velocity evolution equation for $v_{3}$ with $%
1/a^{2}=1/b^{2}.$In the approximation $a\gg b \gg c$ and $b^{4}>a^{2}c^{2}$
for large $t$ from (\eqref{eq:vel1} and \eqref{eq:vel2}), the evolution
equations for $v_{1}$ and $v_{2}$ reduce to:

\begin{equation*}
\dot{v}_{1}+\frac{v_{2}v_{3}}{v_{0}c^{2}}\left( 1-\frac{2L^{3}}{w^{1/2}}%
\right) =0,
\end{equation*}

\begin{equation*}
\dot{v}_{2}-\frac{v_{1}v_{3}}{v_{0}c^{2}}\left( 1+\frac{2L^{3}b^{2}}{%
a^{2}w^{1/2}}\right) =0.
\end{equation*}%
We assume non-relativistic velocities, so take $v_{0}=1$, and note that $%
w=\rho _{r}+p_{r}=4\rho _{r}/3$. Since $\rho _{r}\propto (abc)^{-4/3}\propto
t^{-2}$, we write

\begin{equation*}
w^{1/2}=\frac{M}{t},
\end{equation*}%
where $M$ is a positive constant. Therefore the radiation entropy, $s$,
depends on $M$ via

\begin{equation*}
s\propto \rho _{r}^{3/4}\propto w^{3/4}\propto M^{3/2}.
\end{equation*}

Hence, we have approximately

\begin{equation}
\dot{v}_{1}+\frac{v_{2}v_{3}}{c_{0}^{2}t\ln ^{2}(t)}\left( 1-\frac{2L^{3}t}{M%
}\right) =0,  \label{a}
\end{equation}

\begin{equation}
\dot{v}_{2}-\frac{v_{1}v_{3}}{c_{0}^{2}t\ln ^{2}(t)}\left( 1-\frac{%
2L^{3}b_{0}^{2}t\ }{Ma_{0}^{2}\ }\right) =0,  \label{b}
\end{equation}%
where $v_{3}$ is constant. At large times these equations are (and scaling $%
a_{0}=b_{0})$

\begin{eqnarray}
\dot{v}_{1} &=&\frac{2L^{3}v_{3}v_{2}}{Mc_{0}^{2}\ln ^{2}(t)}\equiv \frac{%
Dv_{2}}{\ln ^{2}(t)},  \label{v1} \\
\dot{v}_{2} &=&-\frac{2L^{3}v_{3}b_{0}^{2}v_{1}}{Mc_{0}^{2}a_{0}^{2}\ \ln
^{2}(t)}\equiv -\frac{Dv_{2}}{\ln ^{2}(t)},  \label{v2}
\end{eqnarray}%
where

\begin{equation*}
D=\frac{2L^{3}v_{3}}{Mc_{0}^{2}}
\end{equation*}%
is a constant. Hence,we see immediately that

\begin{equation}
v_{1}^{2}+v_{2}^{2}=E\mathrm{\ \ :}\text{ }E=\mathrm{\ constant.}
\label{int}
\end{equation}%
Since $v_{1}=\sqrt{E-v_{2}^{2},}$we have in (\ref{v1})

\begin{equation*}
\dot{v}_{1}=-v_{2}\dot{v}_{2}(E-v_{2}^{2})^{-1/2}=\frac{Dv_{2}}{\ln ^{2}(t)},
\end{equation*}%
hence

\begin{equation*}
\int \frac{dv_{2}}{\sqrt{E-v_{2}^{2}}}=-D\int \frac{dt}{\ln ^{2}(t)}
\end{equation*}%
Therefore,

\begin{equation*}
v_{2}=\sqrt{E}\sin \left( -D\int \frac{dt}{\ln ^{2}(t)}\right) ,
\end{equation*}

and so, by (\ref{int}), we have

\begin{equation*}
v_{1}=\sqrt{E}\cos \left( -D\int \frac{dt}{\ln ^{2}(t)}\right) .
\end{equation*}%
The components $v_{1}$ and $v_{2}$ therefore undergo bounded oscillations
while $v_{3}$ remains constant.

We can get a better approx by keeping all the terms in (\ref{a}) and (\ref{b}%
). If we write them as

\begin{equation}
\dot{v}_{1}+\frac{Av_{2}\ }{t\ln ^{2}(t)}\left( 1-Bt\right) =0,
\end{equation}

\begin{equation}
\dot{v}_{2}-\frac{Av_{1}\ }{t\ln ^{2}(t)}\left( 1-Bt\right) =0,
\end{equation}

then $v_{1}^{2}+v_{2}^{2}=$ $E$, and hence we find a second order correction
which confirms the oscillatory behaviour of he velocities with growing
periods of oscillation:

\begin{eqnarray*}
v_{1} &=&E^{1/2}\cos \left( -\frac{1}{\ln (t)}-B\int \frac{dt}{\ln ^{2}(t)}%
\right) \\
v_{2} &=&E^{1/2}\sin \left( -\frac{1}{\ln (t)}-B\int \frac{dt}{\ln ^{2}(t)}%
\right)
\end{eqnarray*}

\begin{acknowledgements}
J.D.B.is supported by the Science and Technology
Facilities Council (STFC) of the United Kingdom. C.G. is supported by the
Jawaharlal Nehru Memorial Trust Cambridge International Scholarship. C.G. would also like to thank Bogdan V. Ganchev for useful discussions.
\end{acknowledgements}

\bibliographystyle{unsrt}
\bibliography{MovingMatterBib}

\begin{thebibliography}{10}

\bibitem{tol}
Richard~C Tolman.
\newblock On the theoretical requirements for a periodic behaviour of the
  universe.
\newblock {\em Physical Review}, 38(9):1758, 1931.

\bibitem{znov}
Iakov~Borisovich Zeldovich and Igorʹ~Dmitrievich Novikov.
\newblock {\em Relativistic astrophysics, 2: The structure and evolution of the
  Universe}, volume~2.
\newblock University of Chicago Press, 1971.

\bibitem{bdab}
John~D Barrow and Mariusz~P Dabrowski.
\newblock Oscillating universes.
\newblock {\em Monthly Notices of the Royal Astronomical Society},
  275(3):850--862, 1995.

\bibitem{me}
John~D Barrow and Chandrima Ganguly.
\newblock Cyclic mixmaster universes.
\newblock {\em Physical Review D}, 95(8):083515, 2017.

\bibitem{matzner}
Richard~A Matzner, LC~Shepley, and James~B Warren.
\newblock Dynamics of so (3, r)-homogeneous cosmologies.
\newblock {\em Annals of Physics}, 57(2):401--460, 1970.

\bibitem{kingellis}
Andrew~R King and George~FR Ellis.
\newblock Tilted homogeneous cosmological models.
\newblock {\em Communications in Mathematical Physics}, 31(3):209--242, 1973.

\bibitem{LukashVII}
VN~Lukash.
\newblock Homogeneous cosmological models with gravitational waves and
  rotation.
\newblock {\em JETP Letters}, 19:265--267, 1974.

\bibitem{Lukashrev}
VN~Lukash.
\newblock Physical interpretation of homogeneous cosmological models.
\newblock {\em Il Nuovo Cimento B (1971-1996)}, 35(2):268--292, 1976.

\bibitem{lukashnovikov}
VN~Lukash, ID~Novikov, and AA~Starobinskii.
\newblock Particle creation in the vortex cosmological model.
\newblock {\em Zhurnal Eksperimentalnoi i Teoreticheskoi Fiziki},
  69:1484--1500, 1976.

\bibitem{lukashstarob}
VN~Lukash, ID~Novikov, AA~Starobinsky, and Ya~B Zeldovich.
\newblock Quantum effects and evolution of cosmological models.
\newblock {\em Il Nuovo Cimento B (1971-1996)}, 35(2):293--307, 1976.

\bibitem{lukash}
LP~Grishchuk, AG~Doroshkevich, and VN~Lukash.
\newblock The model of ``mixmaster universe" with arbitrarily moving matter.
\newblock {\em Soviet Journal of Experimental and Theoretical Physics}, 34:1,
  1972.

\bibitem{tsag}
John~D Barrow and Christos~G Tsagas.
\newblock On the stability of static ghost cosmologies.
\newblock {\em Classical and Quantum Gravity}, 26(19):195003, 2009.

\bibitem{kbm}
John~D Barrow, Dagny Kimberly, and Joao Magueijo.
\newblock Bouncing universes with varying constants.
\newblock {\em Classical and Quantum Gravity}, 21(18):4289, 2004.

\bibitem{mis}
Charles~W Misner.
\newblock Mixmaster universe.
\newblock {\em Physical Review Letters}, 22(20):1071, 1969.

\bibitem{HeckShuk}
O~Heckmann and E~Schucking.
\newblock Relativistic cosmology.
\newblock {\em Gravitation, an Introduction to Current Research}, page 438,
  1962.

\bibitem{chern}
David~F. Chernoff and John~D. Barrow.
\newblock Chaos in the mixmaster universe.
\newblock {\em Phys. Rev. Lett.}, 50:134--137, Jan 1983.

\bibitem{landau}
Lev~Davidovich Landau and Evgenii~Mikhailovich Lifshitz.
\newblock The classical theory of fields.
\newblock 1971.

\bibitem{skew}
John~D. Barrow and Mariusz~P. Da\ifmmode~\mbox{\c{}}\else \c{}\fi{}browski.
\newblock Kantowski-sachs string cosmologies.
\newblock {\em Phys. Rev. D}, 55:630--638, Jan 1997.

\bibitem{star}
AA~Starobinskii.
\newblock Isotropization of arbitrary cosmological expansion given an effective
  cosmological constant.
\newblock {\em JETP lett}, 37(1), 1983.

\bibitem{bkl}
Vladimir~A Belinskii, Isaak~M Khalatnikov, and Evgeny~M Lifshitz.
\newblock A general solution of the einstein equations with a time singularity.
\newblock {\em Advances in Physics}, 31(6):639--667, 1982.

\bibitem{JBIX}
John~D. Barrow.
\newblock Chaos in the einstein equations.
\newblock {\em Phys. Rev. Lett.}, 46:963--966, Apr 1981.

\bibitem{DLN2}
AG~Doroshkevich, VN~Lukash, and ID~Novikov.
\newblock The isotropization of homogeneous cosmological models.
\newblock {\em Sov. Phys.--JETP}, 37:739--46, 1973.

\bibitem{JB}
John~D Barrow.
\newblock On the origin of cosmic turbulence.
\newblock {\em Monthly Notices of the Royal Astronomical Society},
  179(1):47P--49P, 1977.

\bibitem{nohair1}
Robert~M. Wald.
\newblock Asymptotic behavior of homogeneous cosmological models in the
  presence of a positive cosmological constant.
\newblock {\em Phys. Rev. D}, 28:2118--2120, Oct 1983.

\bibitem{nohair2}
John~D Barrow.
\newblock The very early universe.
\newblock In {\em Proceedings of the Nuffield Workshop), ed. GW Gibbons, SW
  Hawking and STC Siklos}, volume 267, 1983.

\bibitem{tip}
F.J. Tipler.
\newblock {Singularities in universes with negative cosmological constant}.
\newblock {\em \apj}, 209:12--15, October 1976.

\end{thebibliography}

\end{document}